\documentclass[12pt]{article}
\usepackage[authoryear]{natbib}
\usepackage[utf8]{inputenc}
\setcitestyle{round}
\usepackage{amsfonts, animate,subfigure, comment}
\usepackage{amssymb,amsthm,dsfont}
\usepackage[margin=1in]{geometry}
\usepackage{blkarray, bbm}
\usepackage{setspace}
\usepackage{amsmath}
\usepackage[format=hang,labelfont=bf,font={small,it}]{caption}
\usepackage[capposition=below]{floatrow}
\usepackage{url}
\usepackage{graphics}
\usepackage{graphicx,color}
\usepackage[colorlinks, urlcolor = blue, citecolor = blue]{hyperref}
\usepackage[ruled]{algorithm2e}
\usepackage{bm}
\usepackage[format=hang,labelfont=bf,font={small,it}]{caption}
\usepackage[capposition=below]{floatrow}
\usepackage{paralist}
\usepackage{soul}

\usepackage{color}

\usepackage[rightbars, color]{changebar}
\cbcolor{blue}
\newcommand{\beq}{\begin{equation}}
\newcommand{\eeq}{\end{equation}}

\newtheorem{prop}{Proposition}
\newtheoremstyle{case}{}{}{}{}{}{:}{ }{}
\theoremstyle{case}
\newtheorem{case}{Case}

\begin{document}
\renewcommand{\thefootnote}{\arabic{footnote}}
\vspace{-3cm}
\title{
    \fontencoding{T1}
  \fontfamily{timesnewroman}
  \fontseries{m}
  \fontshape{it}
  \fontsize{22}{22}
 \selectfont Estimating spillovers using imprecisely measured networks \\ 
 }
\normalfont 
\vspace{1cm}
\author{Morgan Hardy \\
\vspace{.1cm}
New York University - Abu Dhabi \and
Rachel M. Heath\\ \vspace{.1cm}University of Washington \and
Wesley Lee \\
\vspace{.1cm}
University of Washington \and
Tyler H. McCormick\thanks{Contact email: tylermc@uw.edu.  We thank Eric Auerbach, Jing Cai, Arun Chandrasekhar, Jie Gao, Alan Griffith, Jake Murray, Michael Sobel and audience members at the Conference in Development Economics in honor of Chris Udry, the Minghui Yu Memorial Conference at Columbia University, the Frontiers of Networks Science Workshop at New York University Abu Dhabi, and the Causal Learning with Interactions Workshop sponsored by The Alan Turing Institute, CeMMAP, and ERC for helpful suggestions.} \\
\vspace{.1cm}
University of Washington}
%\date{\today}
\date{}
\maketitle
\vspace{-.4cm}

\begin{abstract}
 In many experimental contexts, whether and how network interactions impact outcomes of both treated and untreated individuals are key concerns. Networks data is often assumed to perfectly represent the set of individuals who might be affected by these interactions. This paper considers the problem of estimating treatment effects when measured connections are, instead, a noisy representation of the true spillover pathways. We show that existing methods yield biased estimators in the presence of this mismeasurement error. We develop a new method that uses a class of mixture models to model the underlying network and account for missing connections, and then discuss its estimation via the Expectation-Maximization algorithm. We check our method's performance by simulating experiments on network data from 43 villages in India \citep{banerjee2013diffusion}. Finally, we use data from \citet{cai2015social} to show that estimates using our method are more robust to the choice of network measure than existing methods. 
 \end{abstract}

\onehalfspacing

\section{Introduction} \label{section:Introduction}
Interactions between peers are of interest in many economic settings, such as health \citep{miguel2004worms,Oster2011, godlonton2012peer,anukriti2022bring}, education \citep{angelucci2010family, duflo2011peer, wantchekon2014education}, consumption \citep{de2020consumption}, job search \citep{Magruder2010, Wang2013, Heath2018}, migration \citep{munshi2003networks}, personal finance \citep{bursztyn2014understanding}, politics \citep{cruz2017politician}, agriculture \citep{cai2015social,vasilaky2018good, benyishay2019social, beaman2021can}, and firms  \citep{fafchamps2016networks, cai2018interfirm, hardy2021takes}.  Moreover, even when spillovers to non-treated peers are not of direct interest, the possibility of treatment spillovers to the control group violates the stable unit treatment value assumption (SUTVA) needed to identify causal treatment effects \citep{rubin1974estimating}.  In both cases, knowing the group of peers who are potentially affected by a treatment 
allows researchers to accurately estimate peer effects and assess potential SUTVA violations.

However, measuring social networks is challenging.  It is expensive to collect data on an individual's entire social network \citep{breza2020using}, leading researchers to use data on proxies for networks such as geography \citep{foster1995learning, miguel2004worms, bayer2008place, godlonton2012peer}, familial relationship \citep{Magruder2010, Wang2013, Heath2018} or sharing a common nationality \citep{beaman2011social}, language \citep{bandiera2009social, bandiera2010social}, ethnic group \citep{fafchamps2003ethnicity}, religion \citep{munshi2006social}, or caste \citep{munshi2006traditional, munshi2016networks}.  Even if researchers collect network data,\footnote{There is a substantial empirical literature that addresses the reliability of network data elicited through surveys, with emphasis on the type and salience of relationships being surveyed, temporal dependence, and how links are elicited. See \citet{bell2007partner, marsden2016survey} for reviews.} the set of individuals potentially affected by a given treatment may not entirely correspond to the elicited network if networks are truncated due to concerns about survey fatigue\footnote{\citet{griffith2017how} shows how limiting the number of peers a subject can report in the data from the National Adolescent Health Project can bias estimates in the linear-in-means model.}, or the experiment changes the network itself \citep{comola2017treatment, stein2018re, banerjee2021changes}.  It is also difficult to ask respondents to specify the precise set of individuals potentially affected by a given treatment by asking about either past interactions or hypothetical future interactions \citep{hardy2021takes}.

 In this paper, we focus on experimental settings where the observed network represents a mismeasured version of the true network of treatment interference, allowing for both unreported spillover pathways and misreported links over which no spillovers could occur. We use a local network exposure approach  \citep{ugander2013graph, aronow2017estimating} that defines a set of individuals whose treatment status can potentially affect each subject. \footnote{\cite{chandrasekhar2023general} suggest an extension to the exposure condition framework that allows for arbitrary and continuous dependence, but assumes the network is observed without error.} Each individual is assigned to an ``exposure condition'' which is defined based on the individual's treatment status and the treatment status of their neighbors.  We first show missing links and misreported links in the network can cause mismeasured treatment exposure conditions and hence biased estimators. We develop a class of mixture models that can model the distribution of the latent true exposure conditions and discuss parameter estimation using the Expectation-Maximization (EM) algorithm. These models rely on parametric assumptions about the distribution of missing links conditional on the observed network data as well as parametric assumptions on the behavior of outcomes within each treatment exposure condition. \footnote{In the context of experiments, mixture modeling has previously been used under the potential outcomes framework to address subject compliance \citep{sobel2012compliance}. Subjects are classified into various conditions based on their behavior with respect to treatment assignment (e.g. never takes treatment, complies with treatment, always takes treatment), with the goal of measuring a treatment effect solely for complier subjects. However, this classification is inherently unknown since the behavior of each subject is only observed under a single treatment assignment, and thus estimation proceeds by jointly modeling the uncertainty over these compliance conditions with the treatment outcome under each compliance condition.} Under a linear regression model for the latter assumptions (namely, that the mean outcome within each exposure group is a linear function of the number of treated peers), we prove the mixture model is identifiable and the maximum likelihood estimator from the EM algorithm is consistent. 
 
 We evaluate our method with both simulations and replication of an existing study. We simulate experiments on networks of Indian households from 47 villages \citep{banerjee2013diffusion}. We are able to recover accurate estimates of direct and indirect treatment effects when commonly used Horvitz-Thompson estimators based on weighted averages of outcomes by group fail. Finally, we implement our method using network data from a randomized evaluation of insurance information sessions with rural farmers in China  \citep{cai2015social}. We find that our method produces more consistent estimates of direct and indirect treatment effects, across various choices of network measures, than naive treatment effects estimates that assume the network is measured perfectly.  Code to replicate the results here is available at \url{https://github.com/thmccormick/spillovers-mismeasured-graphs}.

Our results are relevant to many experimental contexts where a subject's behavior or outcome may be influenced by other subjects' treatment assignment in addition to their own.  A common approach in these cases is to randomize treatment at a geographic or organizational level that plausibly contains each treated individual's network of potential spillovers, such as a village (in isolated, rural settings), and then compare treated individuals to ``pure controls'' in non-treated units.  However, even if this is possible, comparing treated to control subjects still confounds treatment effects and spillovers on these treated subjects.\footnote{An exception would be if the treated individuals are a small enough fraction of treated units that they are unlikely to know treated subjects.  Comparing treatment to control individuals would then identify the average direct treatment effect by construction. However, this would likely require a large enough number of units to be impractical or prohibitively expensive in many settings. Treatment effects in such contexts are also not particularly informative about the results from scaling up a treatment to an entire population.}  Moreover, in other cases, a pure control is not feasible, because the experiment must be implemented within a single firm \citep{bandiera2009social, bloom2014does, adhvaryu2016management} or market \citep{conlon2013efficiency}, or it is not possible to leave a large enough buffer between treated and control areas to render spillovers unimportant.  Potential SUTVA violations could then introduce both upward and downward bias in direct treatment effect estimates.\footnote{The reduction in exposure to disease from directly treated school children in Kenya may indirectly improve the health outcomes of school children who did not directly receive the treatment, biasing downward naively estimated benefits of deworming pills \citep{miguel2004worms}. In contrast, increased police patrolling on the streets of Bogota, Colombia, may merely push crime ``around the corner'', biasing upward the estimated impact on crime rates \citep{blattman2021place}.}

In terms of related literature, there are two areas that deserve particular attention: (i) work on mismeasurement/sampling in networks and (ii) models for characterizing peer influence.  Beginning with the first item, our approach is related to a growing literature in economics, political science, sociology, and statistics on network sampling and mismeasurement. This literature views the observed graph as a mismeasured version of a true, unobserved, network.  One common setting assumes that the researcher can perfectly observe a fraction of the total network.   For example, \citet{chandrasekhar2011econometrics} shows how egocentrically sampled network data can be used
to predict the ``full" network in a process they term graphical reconstruction. \footnote{See \citet{williams2016economic} and \citet{griffith2017random} for sample applications.}  By contrast, we study a setting in which all potential links are measured, but may contain some error.  As in \citet{handcock2010modeling} and \citet{newman2018network}, we relate the observed and latent network via a probabilistic model and, given a set of model parameters, construct a distribution over the true network conditional on the observed graph.\footnote{In our setting, by contrast, even a full graph cannot be used to train probabilistic models, because of the potential for error on every link (and non-link). This creates an inability to learn the parameters of the mismeasurement process. For example, the observed network data does not inform the proportion of true links missing from the observed graph.}
 The functional form for network effects we assume in the local network exposure approach \citep{aronow2017estimating, ugander2013graph} generalizes another commonly used approach for measuring spillover effects: a model in which a subject's outcome is modeled as a function of the number of treated peers, conditional on the total number of peers \citep{miguel2004worms}.\footnote{Our local network exposure approach (as well as the ``linear-in-treated-peers'' approach just described) contrasts with linear-in-means models \citep{manski1993identification, bramoulle2009identification} in its assumed avenues of treatment interference. Local network exposure models assume that the avenues of interference for each subject are limited to the treatment assignments of other subjects in their network. On the other hand, linear-in-means models \citep{manski1993identification, bramoulle2009identification} postulate that indirect treatment effects are a linear function of the average outcome of that subject's peers.}  The ``linear-in-treated-peers'' approach typically assumes the (expected) outcome changes as a linear function of the number of treated contacts.  Equivalently, it assumes that the additional effect of each treated contact on the outcome is constant.  The exposure condition setup, in contrast, allows for an arbitrary relationship between the number of treated contacts and the outcome, depending on how exposure conditions are defined.  For the purpose of exposition, we assume a conceptually simple but nonlinear relationship in our data example and simulations.  Specifically, we assume that the effect of knowing treated individuals is constant after the first treated contact.  This setup has the advantage of needing only four exposure conditions (zero treated contacts or more than zero, crossed by each respondent's treatment status). This scenario corresponds to a situation where one treated peer is very influential, say, if a trusted peer conveys reliable information that is not particularly context-specific, such as the existence of a particular initiative. 
 
 Adding more exposure conditions, however, allows us to capture the constant increase implied by the ``linear-in-treated-peers'' model or, in the extreme, to use a totally unstructured approach where we estimate the mean outcome for each number of treated peers separately.  The latter approach would have the advantage of capturing non-monotonic effects. Such monotonicity is relevant, for instance, in a Bayesian learning model in which the first treated peers are the most impactful, but other treated peers contribute less to further Bayesian updating of a respondent's prior if each new peer conveys limited additional information.  
 Of course, this added flexibility comes with additional data requirements, and we see the standard statistical trade-off between parametric assumptions and degrees of freedom.  In Section \ref{section:linear in sums}, we expand this discussion by laying out what happens when we increase the number of exposure conditions, showing specifically the parallel to the ``linear-in-treated-peers'' approach. Additionally, we show simulation and data results using the ``linear-in-treated-peers'' formulation in the Appendix~\ref{sec:LIS}.

This paper proceeds as follows. In the next paragraph, we introduce notation that we will use throughout.  Then, in Section \ref{section:Existing Methods} we discuss existing methods for estimating direct and indirect treatment effects.  In Section \ref{section:AS model mismeasurement}, we derive formulas for the bias in Horvath-Thomson estimators based on weighted averages when networks are measured with error.  In Section \ref{section:Model} we propose a mixture model to estimate treatment effects that can account for latent ties between subjects. We discuss when this model is identified, how to estimate model parameters and treatment effects, and examine model performance using simulations. We apply our methodology to \citet{cai2015social} in Section \ref{section:Application}, and conclude in Section \ref{section:Discussion} with a discussion.

We now introduce some basic notation that we will use throughout the rest of the paper. Let $i \in \{1,2,\cdots, N\}$ index the subjects in the study, with corresponding observed outcomes $y_i$, which we vectorize as ${\bf y}$. For simplicity, suppose treatment is binary with levels ``treatment" (1) and ``control" (0), and the treatment assignment mechanism is random and explicitly known. Denote the vector of treatment assignment with ${\bf t}  \in \{0,1\}^N$, in which the treatment of individual $i$ is $t_i$. Suppose the true influence network $G$ is directed and binary, with the edge $i\rightarrow j$, representing individual $i$'s influence on individual $j$, encoded by $G_{ij} = 1$. Let $G_j$ denote the $j$th column of $G$, indicating the influencers of individual $j$, so $\mathbf{1}^\prime G_j$ is the the number of influencers or in-degree of $j$.\footnote{Analogously, $G_j \cdot \mathbf{1}$ is the number of people that individual $j$ influences, or the out-degree.} For now, let us assume $G$ is observed. Finally, let $\overline{G_j}$ denote the $j$th column of G normalized to sum to 1 ($\mathbf{1}^\prime \overline{G_j} = 1 $) unless $\mathbf{1}^\prime G_j = 0$, in which case $\overline{G_j} = G_j = \mathbf{0}$.

\section{Measuring network spillovers in experiments} \label{section:Existing Methods}

\subsection{Local network exposure model} \label{section:AS Model}

\citet{aronow2017estimating} and \citet{ugander2013graph} propose estimators for average direct and indirect treatment effects by building on the Rubin causal model \citep{rubin1974estimating}. In the context of experiments, each subject has a set of ``potential outcomes" $(Y_i(t_i = 0), Y_i(t_i = 1))$ corresponding to the possible outcomes under each treatment (or none). The inference task is to estimate
the average treatment effect, defined to be the difference between the average outcome of the population if the entire population was treated and the average outcome if the entire population was in the control:
\begin{equation}
ATE(1,0) = \frac{1}{N}\sum_{i=1}^{N} \left[Y_i(t_i = 1) - Y_i(t_i = 0)\right].
\label{eq:ATE}
\end{equation}
This quantity is not observed since we cannot observe the full set of potential outcomes for each subject, but assuming completely random assignment can be estimated by the difference in sample means:
\begin{equation}
\widehat{ATE(1,0)} = \frac{1}{N_1}\sum_{i=1}^{N} y_i \mathbf{1}[t_i = 1] - \frac{1}{N_0}\sum_{i=1}^{N} y_i \mathbf{1}[t_i = 0],
\label{eq:ATE_est}
\end{equation}
where $N_k$ is the number of subjects in treatment $k$. A crucial assumption in the Rubin causal model is SUTVA, which states than a subject's potential outcomes are unaffected by the treatments of other subjects. In experiments on networks, SUTVA is violated if the treatments of peers influence the outcomes for an individual.

\citet{aronow2017estimating} considers a violation of SUTVA by allowing for individuals to be systematically affected by the treatment assignments of their peers. By making assumptions that restrict the nature of these influences, they induce mappings of the treatment vector ${\bf t}$ to distinct ``exposure conditions", or what \citet{manski2013identification} terms ``effective treatments." In a simple instance of their framework, which we borrow for our model in Section \ref{section:Model}, individuals are affected by whether or not any of their influencers in $G$ are treated, inducing a random assignment into one of four exposure conditions, corresponding to levels of direct and indirect exposure to treatment:

\begin{equation}
C_i \equiv C_{i}(t,G_i) = \left\{\begin{array}{llr}
     c_{00} & \text{(No Exposure)}: & t_i = 0 \text{ and } {\bf t}^\prime G_i = 0\\
     c_{01} & \text{(Indirect Exposure)}: & t_i = 0 \text{ and } {\bf t}^\prime G_i > 0\\
     c_{10} & \text{(Direct Exposure)}: & t_i = 1 \text{ and } {\bf t}^\prime G_i = 0\\
     c_{11} & \text{(Full Exposure)}: & t_i = 1 \text{ and } {\bf t}^\prime G_i > 0.
\end{array}\right.
\label{eq:exposure conditions}
\end{equation}
In this model, both direct and indirect effects are taken to be binary, with an individual being indirectly exposed to treatment if one or more of their influencers are (directly) treated. Each subject $i$ would have four potential outcomes $(Y_i(C_i = c_{00}),Y_i(C_i = c_{01}),Y_i(C_i = c_{10}),Y_i(C_i = c_{11}))$, one for each exposure condition. Note this setup assumes that the number of connections treated does not have an effect beyond the presence or absence of at least one, and an individual can only be influenced by a first-order connection in the network. 

The choice of indirect exposure can be related to diffusion models of information and disease in which ``contagion" can occur given a single source of exposure \citep{centola2007complex}, also called simple contagion models.  In a Bayesian learning framework, these models would be relevant in cases where individuals do not have strong priors (so that the first piece of information they receive is the most important) and where the information is non-rival and relatively costless to pass along (so that a treated network member would be very likely to pass on information).  Moreover, rational individuals in a Bayesian learning context infer that information shared by multiple network members is likely come from a common source, so additional information will be less valuable.\footnote{Alternatively, other social learning models assume that individuals are boundedly rational and do not infer that information shared by multiple network members likely comes from a common source \citep{degroot1974reaching,banerjee2019naive}.}.  By contrast, in settings in which individuals have a stronger prior -- which is different from the information they receive -- or information is costly to pass along, the number of treated peers matters; these settings are sometimes called complex contagion models.  For instance, an individual adopts a technology if a fraction of her network that is above some threshold adopts the technology \citep{granovetter1978threshold, acemoglu2011diffusion, beaman2018can}.  In such cases, the assumption that only the first treated peer matters can be relaxed by adding additional exposure conditions that correspond to the appropriate model of peer influence in a given context.\footnote{For instance, \citet{beaman2018can} find evidence in favor of a threshold model in which at least two treated peers is necessary for adoption of a new technology.}  Similarly, the assumption that only first-order links matter could be relaxed by adding exposure conditions corresponding to second-order exposure.

The primary quantities of interest would then be given by average treatment effects akin to equation (\ref{eq:ATE}):
\begin{equation}
ATE(c,d) = \frac{1}{N}\sum_{i=1}^{N} \left[Y_i(C_i = c) - Y_i(C_i = d)\right].
\label{eq:ATE2}
\end{equation}
The average direct treatment effect would be given by $ATE(c_{10},c_{00})$, while the average indirect treatment effect when not directly treated would be $ATE(c_{01},c_{00})$. Estimating these quantities is equivalent to estimating the mean outcomes of the entire population under each exposure condition:
\begin{equation}
\mu_c = \frac{1}{N}\sum_{i=1}^{N} Y_i(C_i = c),
\label{eq:mu}
\end{equation}
so we focus on the latter for this section and the next, with the additional assumption that each subject is assigned to treatment with some constant and known probability independently of other subjects. Note if certain subjects have zero probability of being placed in certain exposure conditions, e.g. when a subject has no influencers, estimation must be restricted to the sub-population of individuals with non-zero probability of being placed in every condition. In contrast to the case when the SUTVA assumption is satisfied, we cannot estimate these means using just their sample counterparts. Variability in the in-degrees of individuals causes variation in the probabilities of assignment to each exposure condition. Namely, individuals with high in-degree are more likely to be indirectly exposed to the treatment since they have more influencers who potentially may be treated. This selection bias could affect the mean estimates if there is heterogeneity in the outcomes within exposure conditions associated with in-degree. Horvitz-Thompson estimators use inverse probability weighting in order to take varying exposure probabilities into account to produce unbiased estimators of these mean outcomes:
\begin{equation}
\widehat{\mu}_{c,HT} = \frac{1}{N}\sum_{i=1}^{N} \frac{y_i\cdot\mathbf{ 1}_{[C_i = c]}}{P(C_i = c)}.
\label{eq:HT_est}
\end{equation}
Note this estimator is equivalent to the sample mean if the probability of assignment to an exposure condition is constant among subjects. These estimators are unbiased regardless of the form of the heterogeneity between the outcomes and network degrees.\footnote{However, they can have high variance when the exposure conditions are highly imbalanced on in-degree. This would arise when the probabilities $P(C_i = c)$ are small for some $i$, yielding large weights $\frac{1}{P(C_i = c)}$.  This suggests potential efficiency gains from stratifying on degree.} 

Explicitly modeling the relationship between potential outcomes and network degrees can result in lower variance estimators at the cost of additional assumptions about the validity of these relationships. For example, suppose we believe that for each exposure condition $c$, the relationship between the in-degree $\left(\mathbf{1}^\prime G_i\right)$ and the potential outcome $Y_i(C_i = c)$ can be modeled with
\begin{equation}
Y_i(C_i = c) \sim f\left(\cdot;\theta_c, \mathbf{1}^\prime G_i\right),
\label{eq:potential outcome f model}
\end{equation}
where $\mathbf{1}^\prime G_i$ is the in-degree of individual $i$ and $\theta_c$ are model parameters. Assuming this model accurately characterizes the relationship between the potential outcomes and in-degrees, the distribution of potential outcomes is conditionally independent of the exposure assignment (induced by the treatment assignment) vector given the in-degrees of the subjects, such that the exposure assignment mechanism can be ``ignored" during inference of the means \citep{rubin1974estimating}. The estimate of the mean outcome under exposure condition $c$ would then be given by
\begin{equation}
\widehat{\mu}_{c,R} = \frac{1}{N}\sum_{i=1}^{N} \widehat{Y_i}(C_i = c) = \frac{1}{N}\sum_{i=1}^{N} E_{f\left(\cdot;\widehat{\theta}_c, \mathbf{1}^\prime G_i\right)}[y_i],
\label{eq:f est}
\end{equation}
provided an estimate of model parameters $\widehat{\theta}_c$. Parametric models $f\left(\cdot;\theta_c, \mathbf{1}^\prime G_i\right)$ of the outcomes under condition $c$ and in-degree $d$ are necessary for likelihood-based approaches to estimation and are used in the model we propose in Section \ref{section:Model}. A common model familiar to many economists is 
\begin{equation}
f\left(\cdot;\theta_c \equiv (\alpha_c,\beta_c,\sigma^2), \mathbf{1}^\prime G_i\right) = \text{N}\left(\cdot;\alpha_c + \beta_c \mathbf{1}^\prime G_i, \sigma^2\right),
\label{eq:f linear}
\end{equation}
which corresponds to a linear model with different intercepts and slopes for each exposure condition (but common variance). In this case, the estimates of mean outcome would be given by $\widehat{\mu}_{c,R} = \widehat{\alpha}_c + \widehat{\beta}_c \frac{1}{N}\sum_{i=1}^{N} \mathbf{1}^\prime G_i.$

\subsection{The linear-in-treated-peers approach} \label{section:linear in sums}
In this section, we discuss the commonly used ``linear-in-treated-peers'' regression approach and how it is encapsulated in the exposure conditions setup we use.  As mentioned previously, the exposure conditions framework generalizes another popular approach to accounting for and measuring treatment spillovers: the linear-in-treated-peers model \citep{miguel2004worms} in which treatment is a linear function of the number of treated peers.  Specifically,
\[y_i=\omega_0+\omega_1\sum_j G_{ij}t_j+\omega_2\sum_j G_{ij}+X_{ij}'\omega+\epsilon_i\].

As with any linear model, there is a trade-off in that this model requires a strong assumption on the form of dependence (i.e. constant increase in expected outcome for each additional treated peer), but requires estimating few parameters. In the context where there is imperfect measurement of the graph we have, 

\[y_i=\omega_0+\omega_1\sum_j \widetilde{G}_{ij}t_j+\omega_2\sum_j G_{ij}+X_{ij}'\omega+\epsilon_i\].

The exposure condition framework in this paper captures the structure of the ``linear-in-treated-peers'' model, while also allowing for more complex forms of dependence. To see this, we can relate the exposure condition framework to regression in terms of expected outcomes, simply noting that $E(y_i)=E\left( \sum_c \mathbbm{1}_{i\in c}Y_i(C_i=c)\right)$, where $\mathbbm{1}$ is the indicator variable. That is, the expected outcome for person $i$ (not conditioned on the exposure condition) is the sum over the expected outcome conditional on the exposure condition, multiplied by the exposure condition indicator. Using the exposure-specific means defined in the previous section, we have $E(y_i|\mu_{c,R})=E\left( \sum_c \mathbbm{1}_{i\in c}\text{N}\left(\cdot;\alpha_c + \beta_c \mathbf{1}^\prime G_i, \sigma^2\right)\right)$.

Conceptually, this framework allows for an arbitrary dependence between the number of treated individuals and the outcome, accounting for variation in total network size.  To see this, we simply define an exposure condition for each possible number of treated connections, up to the maximum observed degree in the graph.  Then, allow $\alpha_c$ to be completely unrestricted between exposure conditions.  In practice, of course, this strategy will yield an extremely high level of uncertainty, and some structure on the exposure condition specific parameters, $\alpha_c$, is appealing.  The ``linear-in-treated-peers'' approach is one option that assumes that the rate of increase in the expected outcome is the same for each additional treated friend.  

To formally define the ``linear-in-treated-peers'' model, we can define the following exposure condition means.  
For exposure conditions 1 and 2 (with no treated friends):
$$f\left(\cdot;\theta_c \equiv (\alpha_c,\beta_c,\sigma^2), \mathbf{1}^\prime G_i\right) = \text{N}\left(\cdot; \mathbbm{1}_{treated}\gamma + \beta_1 \mathbf{1}^\prime G_i+ X_{ij}'\beta, \sigma^2\right)
$$

For exposure conditions 3 and 4 (with one treated friend):
$$f\left(\cdot;\theta_c \equiv (\alpha_c,\beta_c,\sigma^2), \mathbf{1}^\prime G_i\right) = \text{N}\left(\cdot; \eta+ \mathbbm{1}_{treated}\gamma + \beta_1 \mathbf{1}^\prime G_i+X_{ij}'\beta, \sigma^2\right)
$$

For all the rest of the exposure conditions we will say that there are $k$ treated friends:
$$f\left(\cdot;\theta_c \equiv (\alpha_c,\beta_c,\sigma^2), \mathbf{1}^\prime G_i\right) = \text{N}\left(\cdot; k\eta+ \mathbbm{1}_{treated}\gamma + \beta_1 \mathbf{1}^\prime G_i+X_{ij}'\beta, \sigma^2\right)
$$

Said another way, exposure conditions that pertain to one treated peer have an expected mean shift equal to $\eta$ and the exposure condition corresponding to $k$ treated peers expects a mean shift equal to $k\eta$. When we look at the expected value of $y_i|C$, we now have

$$E(y_i|C)=k\alpha+\mathbbm{1}_{treated}\gamma+\beta \mathbf{1}^\prime G_i$$
where having no treated contacts corresponds to $k=0$. We discuss the details of implementing this model in Appendix~\ref{sec:LIS} and also provide results from simulations and observed data.

\section{Characterizing the bias in local network exposure model under mismeasurement} \label{section:AS model mismeasurement}

In this section, we derive the bias in Horvitz-Thompson estimators (\ref{eq:HT_est}) if using a mismeasured network, $\widetilde{G}$, to estimate exposure conditions instead of the true network $G$. We allow $\widetilde{G}$ to be mismeasured such that there are either links present in $\widetilde{G}$ that are not in ${G}$ or vice versa.  Suppose our treatment assignment mechanism ${\bf t}$ is constructed such that each subject $i$ has positive probability of being placed in treatment and positive probability of being placed in control. We can break the impact of using $\widetilde{G}$ in estimation into two distinct factors. First, note that the Horvitz-Thompson estimator can only be used for subjects with non-zero probability of being placed in each exposure condition. Namely, subjects with zero in-degree must be excluded, reflecting the idea that a potential outcome under indirect exposure only makes sense if the subject could be indirectly exposed to treatment under some hypothetical treatment assignment. When we observe a mismeasured version of the network, we may not be able to accurately identify which subjects should be excluded. Certain individuals who have positive in-degree in $G$ may be observed to have zero in-degree in $\widetilde{G}$ and thus would be incorrectly excluded for estimation. At the same time, certain individuals with $\mathbf{1}^\prime G_i=0$ may be observed to have positive in-degree and thus be included during estimation. If either of these situations arose, our estimated average outcomes would then represent a different subpopulation than the true population of subjects with non-zero in-degree.

Second, even if we are able to accurately identify all subjects with non-zero in-degree, bias in mean estimates may be induced by distorted observed exposure conditions. Subjects who are in truth indirectly exposed to treatment would not be observed to be indirectly exposed if all connections to influencers who are treated are unobserved (and no false links to other treated individuals are observed). Similarly, subjects not indirectly exposed to treatment may be falsely observed to be indirectly exposed. The mismeasured exposure conditions are able to correctly identify the level of direct treatment for each subject but not necessarily the level of indirect treatment. Mathematically, observing $\widetilde{C}_i \equiv C_i({\bf t},\widetilde{G}_i) = c_{kl}$ for any $k,l\in\{0,1\}$ may correspond to either $C_i({\bf t},G_i) = c_{k0}$ or $C_i({\bf t},G_i) = c_{k1}$. Recall that in this notation the first subscript denotes the direct treatment condition (whether $i$ is directly treated or not) and the second subscript denotes the indirect treatment (whether at least one member of $i$'s network was treated). The Horvitz-Thompson estimators for each treatment exposure condition $c$ under the mismeasured network $\widetilde{G}$ are given by

\begin{equation}
\widehat{\mu}_{c,HT,\widetilde{G}} = \frac{1}{N}\sum_{i=1}^{N} \frac{y_i\cdot\mathbf{ 1}_{[\widetilde{C}_i = c]}}{P(\widetilde{C}_i = c)}.
\label{eq:HT_est_corr}
\end{equation}
where observed $y_i = \sum_c Y_i(C_i = c)\mathbf{ 1}_{[C_i = c]}$ is dependent on the true exposure condition and the probabilities are taken over possible treatment assignments ${\bf t}$. Holding the observed and true networks fixed and taking the expectation of the estimators $\widehat{\mu}_{c,HT,\widetilde{G}}$ over the possible treatment assignments ${\bf t}$ we have:
\begin{align}
E\left[\widehat{\mu}_{kl,HT,\widetilde{G}}\right] & = \frac{1}{N}\sum_{i=1}^{N} \frac{E\left[y_i\cdot\mathbf{ 1}_{[\widetilde{C}_i = c_{kl}]}\right]}{P(\widetilde{C}_i = c_{kl})} \\
 & = \frac{1}{N}\sum_{i=1}^{N} \frac{ \sum_{c} Y_i(C_i = c) P\left(\widetilde{C}_i = c_{kl}, C_i = c\right)}{P(\widetilde{C}_i = c_{kl})} \\
 & = \frac{1}{N}\sum_{i=1}^{N} Y_i(C_i = c_{k0}) P\left(C_i = c_{k0} | \widetilde{C}_i = c_{kl}\right) + Y_i(C_i = c_{k1}) P\left(C_i = c_{k1}|\widetilde{C}_i = c_{kl}\right).
 \label{eq:HT_bias}
\end{align}
We find the mean estimate of the $c_{kl}$ conditioned on $\widehat{\mu}_{kl,HT,\widetilde{G}}$ under the mismeasured network $\widetilde{G}$ tends to lie between the mean outcomes under the two exposure conditions corresponding to the same level of direct treatment: $\mu_{k0}$ and $\mu_{k1}$.  The bias will be greater with a large probability of mismeasurement 
($P(C_i = c_{k0} | \widetilde{C}_i = c_{kl})$
and $P(C_i = c_{k1}|\widetilde{C}_i = c_{kl})$ are far from 1) and a substantial difference in outcomes between those who are actually indirectly treated versus not ($Y_i(C_i = c_{k0})$ is far from $Y_i(C_i = c_{k1})$). In section \ref{section:Simulations}, we will perform a simulation study that investigates the level of bias in these Horvitz-Thompson estimators and our proposed EM estimates at different rates of unreported true links and falsely observed links.

\section{Latent Variable Model for Network Spillovers} \label{section:Model}

In this section, we propose a latent variable approach to estimating average treatment effects when the network observed is a noisy representation of the true network of interest. 
We assume that each true edge $G_{ij} = 1$ is not observed ($\widetilde{G}_{ij} = 0$) with probability $p$, non edges $G_{ij} = 0$ are observed with probability $q$, and edges are observed/not observed independently of one another. These corruption mechanisms assume the observed edges are a random subset of the true edges and the false edges are a random subset of the non-edges.  However, we can relax this assumption to allow adding/subtracting edges to depend on observed covariates, and do so in the empirical application in section \ref{section:Application}.

\subsection{Latent Variable Model} \label{section:latent model}

Suppose that the true network of interest $G$ is unobserved and we only observe a mismeasured network $\widetilde{G}$. Furthermore, assume the effects of treatment can be characterized with the four exposure conditions defined in (\ref{eq:exposure conditions}). For individual $i$, we observe mismeasured exposure condition $C_i(t,\widetilde{G}_i)$ and in-degree $\mathbf{1}^\prime \widetilde{G}_i$. Assuming known in-degree does not limit the scope of our work, given available methods to consistently estimate degree \citep{mccormick2010many,breza2020using,breza2023consistently} using survey questions that could also be included. The statistical problem is then to model the relationship between these mismeasured statistics and their true, latent, counterparts. Given a distribution over the true exposure condition $C_i(t,G_i)$ and in-degree $\mathbf{1}^\prime G_i$, we can use models like those in equations (\ref{eq:potential outcome f model}) and (\ref{eq:f est}) to estimate mean outcomes for each exposure category. For notational simplicity, let $\widetilde{C}$ represent the vector of mismeasured exposure conditions, $\mathbf{1}^\prime \widetilde{G}$ the vector of mismeasured in-degree, and $C$ and $\mathbf{1}^\prime G$ the corresponding latent terms.

Consider subject $i$, who has exposure condition $C_i(t,G_i) \equiv c_{kl}$, degree $\mathbf{1}^\prime G_i \equiv d$, and $t^\prime G_i \equiv d_t$ connections with treated subjects, but for whom we observe exposure condition $C_i(t,\widetilde{G}_i) \equiv c_{\tilde{k}\tilde{l}}$, degree $\mathbf{1}^\prime \widetilde{G}_i \equiv \tilde{d}$, and $t^\prime \widetilde{G}_i \equiv \tilde{d}_t$ connections with treated individuals instead. Holding the treatment assignments to be fixed, we can separately model the number of connections to treated subjects $d_t$ and the number of connections to not-treated subjects $d - d_t$, from which we can derive the induced exposure conditions. Note this procedure works for any indirect exposure conditions entirely characterized by the number of treated connections and the number of total connections (e.g. ratio of treated connections) and not just (\ref{eq:exposure conditions}).
Following Bayes' rule and noting we observe $\tilde{d}_t$ treated connections when $x$ of the $d_t$ actual treated connections are dropped and another $\tilde{d}_t - d_t + x$ false connections to treated individuals are observed,
\begin{align}
    P(t^\prime G_i = d_t|t,t^\prime \widetilde{G}_i = \tilde{d}_t;p,q) & \propto P(t^\prime \widetilde{G}_i = \tilde{d}_t|t,t^\prime G_i = d_t;p,q) P(t^\prime G_i = d_t) \\
    & \propto \sum_{x=0}^{d_t}\text{Bin}(x;d_t,p)\text{Bin}(\tilde{d}_t - d_t + x;\mathbf{1}^\prime t - t_i - d_t,q) P(t^\prime G_i = d_t)
    \label{eq:num trt friends}
\end{align}
where $\text{Bin}(x;n,p)$ is the probability of $x$ successes from a binomial distribution with $n$ attempts and success probability $p$. Similarly for connections for non-treated subjects,
\begin{equation}
\begin{split}
    P((1-t)^\prime G_i = d_{nt}|t,(1-t)^\prime \widetilde{G}_i = \tilde{d}_{nt};p,q) & \propto \sum_{x=0}^{d_{nt}}\text{Bin}(x;d_{nt},p)
    \\ & \quad \times\text{Bin}(\tilde{d}_{nt} - d_{nt} + x;N - 1 - \mathbf{1}^\prime t + t_i - d_{nt},q) \\ & \quad \times P((1-t)^\prime G_i = d_{nt})
    \label{eq:num non-trt friends}
    \end{split}
\end{equation}
Both sets of equations require a (prior) model over the number of true connections to treated and un-treated subjects. Assuming no additional information about the structure of the true network, one of the most simplistic models would be to model the true graph as an Erdos-Renyi graph, where the probability of a link between any given edges is constant, leading to independence across edges. Under this model, the number of connections to treated/un-treated subjects could be modeled with binomial distributions. However, in many real-world networks we find that the degree distribution demonstrates extra-binomial variation, where differences in degree arise not just from random variation in link formation but also from differences in the propensity to form links.  Thus in the following sections we prefer to use a beta-binomial model.  With a beta-binomial distribution, we can think of each degree as being sampled from a binomial distribution $d\sim\text{Binom}(N-1,p)$, where $p$ is independently sampled from a Beta distribution $p\sim\text{Beta}(\mu,\rho)$, where we paramaterize the beta-binomial distributions in terms of an average probability of success $\mu$ and an overdispersion parameter $\rho$. The variance of this beta-binomial would be given by $(N-1)\mu(1-\mu)\left(1+(N-2)\rho\right)$, compared to $(N-1)\mu(1-\mu)$ for a binomial distribution with parameter $\mu$. We leave the these parameters to be chosen on a application-by-application basis, noting that the choice of these parameters are more influential when there is high mismeasurement in the network and hence higher uncertainty over the true degrees \footnote{Via simulations, we find that a good choice of $\mu$, which governs the overall density of the true network, is more important for our model to recover unbiased estimates.}.

Using the above equations, we can express the relationship between the true exposure condition and degree and their observed counterparts:
\begin{align}
    \tau_i(c_{kl},d;p,q) & \equiv P(C_i(t,G_i) = c_{kl},\mathbf{1}^\prime G_i = d | \widetilde{G}_i,t;p,q) \\
    & = \left\{\begin{array}{lr}
        P(t^\prime G_i = 0, \mathbf{1}^\prime G_i=d |t,\widetilde{G}_i;p,q), & l = 0\\
        P(t^\prime G_i > 0, \mathbf{1}^\prime G_i=d |t,\widetilde{G}_i;p,q), & l = 1
        \end{array}\right.\\
    & = \left\{\begin{array}{lr}
        P(t^\prime G_i = 0|t,\widetilde{G}_i;p,q)P((1-t)^\prime G_i = d|t,\widetilde{G}_i;p,q), & l = 0\\
        \sum_{d_t=1}^{d}P(t^\prime G_i = d_t|t,\widetilde{G}_i;p,q)P((1-t)^\prime G_i = d-d_t|t,\widetilde{G}_i;p,q), & l = 1
        \end{array}\right.
\label{eq:mixture probabilities}
\end{align}

Equation (\ref{eq:mixture probabilities}) defines a distribution over the true, unobserved exposure condition and in-degree, conditional on the treatment vector and the number of observed treated and non-treated connections for individual $i$. When coupled with a parametric model $f\left(\cdot;\theta_c,d\right)$ (see \ref{eq:potential outcome f model}) for the potential outcomes under each (true) exposure condition $c$ and in-degree $d$, we can model the observed outcome $y_i$ as arising from a mixture of the $f\left(\cdot;\theta_c,d\right)$ with weights corresponding to the probabilities $\tau_i(c_{kl},d;p,q)$ over the unobserved quantities (namely, true treatment status $c_{kl}$ and degree $d$).\footnote{One downside of the Horvitz-Thompson estimator (\ref{eq:HT_est}) is that it does not model individual potential outcomes and thus is less amenable to likelihood-based approaches.} The log-likelihood of the parameters $\Theta \equiv \{\theta_{00},\theta_{01},\theta_{10},\theta_{11},p,q\}$ given $y$ is
\begin{equation}
l(\Theta) = \sum_{i=1}^{N}\log\left[ \sum_{c}\sum_{d=0}^{N-1} \tau_i(c,d;p,q)f\left(y_i; \theta_c, d\right) \right].
\label{eq:loglikelihood}
\end{equation}
This is a mixture model in the sense that the likelihood contribution of each subject is the average of her outcome under each exposure condition, weighted by the probability of being in each exposure condition given observed data.  Estimation of the parameters $\Theta$ are provided using maximum likelihood estimation via the Expectation-Maximization algorithm, details of which are provided in Section \ref{section:Estimation}. Note that likelihood estimation is only justified if the observed outcomes are representative of the potential outcomes under each exposure condition, conditional on the true in-degrees. That is only the in-degree can determine indirect exposure to treatment, as in the case of a  random treatment mechanism\footnote{Stratified sampling based on known covariates could also be addressed by directly introducing these covariates into the model.}.

Provided an estimate of the model parameters $\widehat{\Theta}$, estimating the mean outcome under exposure condition $c$ (recall equation (\ref{eq:mu})) is straightforward and given by the expectation of the potential outcome under exposure $c$ for each subject averaged across the population. We estimate $\mu_c$ with the following plug-in estimator:
\begin{align}
\widehat{\mu}_{c} & = \frac{1}{N}\sum_{i=1}^{N} E\left[Y_i(C_i = c) | y,t,\widetilde{G}\right]\\
& = \frac{1}{N}\sum_{i=1}^{N}\sum_{d} P(\mathbf{1}^\prime G=d|y,t,\tilde{G})E_{f\left(\cdot;\widehat{\theta}_c, d\right)}[y_i]\\
& = \frac{1}{N}\sum_{i=1}^{N}\sum_{d} \left(\sum_{c^\prime} \tau_i(c^\prime,d;\widehat{p},\widehat{q})\right)E_{f\left(\cdot;\widehat{\theta}_c, d\right)}[y_i].
\label{eq:mix estimate}
\end{align}

\subsection{Identification} \label{section:Identification}

Before we discuss estimation strategies for our mixture model (\ref{eq:loglikelihood}), we will (partially) characterize the conditions under which this model is identifiable. Without model identifiability, estimation may be unstable and parameters estimates uninterpretable. In this section, we assume $f\left(\cdot;\theta_c,d\right)$ arise from a common univariate family of distributions parameterized by $\eta\equiv\eta(\theta_c,d)$. 

In general, mixture models are trivially unidentifiable since relabeling components yields different parameterizations of a model with the same marginal distribution (see Chapter 1.5 of \citet{mclachlan1988mixture}). In our case, for example, one could relabel direct treatment are indirect treatment and vice versa.  This identifiability issue is of particular concern in our setting, where the labeling of the components is inherently meaningful; for example, being unable to disentangle clusters corresponding to no treatment and indirect treatment would leave us unable to estimate the direction of any indirect treatment effect. We are able to leverage the structure from our mismeasurement model and the linear relationships between mixture components with the same exposure condition to prevent such relabeling from occurring.

Following \citet{fruhwirth2006finite}, we use ``generic identifiability" to refer to identifiability problems not solved by permuting component labels. Generic identifiability holds for mixtures of Gaussians and many other univariate continuous distributions, with the major exceptions being the binomial and uniform distributions. For the binomial distribution, generic identifiability holds if a sufficient number of trials/observations per subject are observed, dependent on the number of components. See \citet{fruhwirth2006finite} for a review of generic identifiability issues.

Note that the fact that the model is not identified for binary outcomes means that it cannot be directly applied to settings with a single, one-time measures of technology adoption.  While this is a limitation of our method given that  technology adoption is an important outcome in the literature on networks, it can be applied to other measures of adoption such as input usage \citep{conley2010learning}, or determinants of adoption such as knowledge about the new technology (as in the example from \citet{cai2015social} in section \ref{section:Application}, or \citet{beaman2018can}).

Unfortunately, generic identifiability of the mixture model (\ref{eq:loglikelihood}) does not directly follow from the generic identifiability of the family $f$, as  \citet{hennning2000identifiability} showed in the case of mixtures of linear regression models. For example, in a mixture of simple linear regressions with two distinct covariate values ${0,1}$ and common variance $\sigma^2$, an equal mixture of $f(x) = x$ and $f(x) = 1-x$ yields the same model as a equal mixture of $f(x) = 0$ and $f(x) = 1$. Observations from a third covariate value would yield generic identifiability. While not immediately applicable to our class of models since in-degree (our covariate) is also latent, \citet{hennning2000identifiability} and \citet{grun2007finite} define conditions under which mixtures of linear and generalized linear models are generically identifiable.

Next, we explicitly prove the identifiability of our mixture model under the regression model (\ref{eq:f linear}) for $f$. Results are readily generalizable to other $f$ that arise from generically identifiable families provided that distinct values of $d$ would allow for the identification of our model parameters $\theta_c$ from the distribution parameters $\eta(\theta_c,d)$.

\begin{prop}
Let $f$ be defined as in (\ref{eq:f linear}) and $\tau_i$ as in (\ref{eq:mixture probabilities}). Assume $p,q,p^\prime,q^\prime<1$\footnote{Both of these edge cases are relatively uninteresting, as when $p=1$ all true edges are not observed and when $q=1$ all non-edges are falsely observed.} and that indirect exposure has some effect (i.e. $\theta_{00} \neq \theta_{01}$ and $\theta_{10} \neq \theta_{11}$). Then
\begin{equation}
\sum_{c}\sum_{d=0}^{N-1} \tau_i(c,d;p,q)f\left(y_i; \theta_c, d\right) = \sum_{c}\sum_{d=0}^{N-1} \tau_i(c,d;p^\prime,q^\prime)f(y_i; \theta_c^\prime, d)
\label{eq:ident setup}
\end{equation}
for all given $t^\prime \widetilde{G}_i = \tilde{d}_{t}$ and $(\mathbf{1}-t)^\prime \widetilde{G}_i = \tilde{d}_{nt}$ implies $\{\theta_{00},\theta_{01},\theta_{10},\theta_{11}\} = \{\theta_{00}^\prime,\theta_{01}^\prime,\theta_{10}^\prime,\theta_{11}^\prime\}$ as long as there exists two distinct $d$ such that we have subjects under each direct treatment status with observed degree $\tilde{d}_t + \tilde{d}_{nt} = d$, and, of these subjects, some  have treated connections $\tilde{d}_t > 0$ while others do not, with $\tilde{d}_t = 0$.
\end{prop}

\begin{proof}
See Appendix \ref{section:IDproof}.
\end{proof}

\subsection{Estimation} \label{section:Estimation}

Maximizing the log-likelihood (\ref{eq:loglikelihood}) with respect to the parameters $\Theta$ cannot be done in closed form due to the summations inside the logarithmic terms. However, if we had directly observed the latent variables $\{C,\mathbf{1}^{\prime}G\}$, the log-likelihood of the parameters $\Theta$ given $y$, $C$ and $\mathbf{1}^\prime G$ would be given by
\begin{equation}
l(\Theta) = \sum_{i=1}^{N}\log\left[\tau_i(c,d;p,q)f\left(y_i; \theta_c, d\right) \right].
\label{eq:complete loglikelihood}
\end{equation}
This would be substantially easier to work with, due to the lack of summation inside the logarithmic terms. Essentially, estimation would entail four regressions, for each exposure condition.  The EM algorithm \citep{demspter1977maximum} is a well-established technique for maximum likelihood estimation in the presence of latent variables that leverages this disparity between the two log-likelihood expressions. Given some set of initial parameter values  $\widehat{\Theta}^0$, the algorithm alternates between estimating posterior distribution of latent variables for each subject given the current parameter values (E-step) and updating the parameter values given these posterior probabilities (M-step). Explicitly working with the latent variables in the M-step yields simpler maximization problems. Each iteration of the algorithm increases the log-likelihood, leading to a local optimum, and the algorithm is run from multiple initialization values in order to maximize the chances of finding a global optimum.

Suppose at iteration $t$ we have parameter estimates $\widehat{\Theta}^{(t)}$. In the E-step, we compute the posterior probabilities over the latent variables using the current parameter estimates. These probabilities, or ``responsibilities," are given by
\begin{align}
\widehat{\gamma}_{icd}^{(t+1)} & \equiv P\left(C_i(t,G_i) = c,\mathbf{1}^\prime G_i = d | y_i, t, \widetilde{G}_i; \widehat{\Theta}^{(t)}\right)\\
& \propto \tau_{i}^{(t)}(c,d;\widehat{p}^{(t)},\widehat{q}^{(t)})f\left(y_i;\widehat{\theta}^{(t)}_c,d\right).
\label{eq:EM responsibilities}
\end{align}
In the M-step, we use these responsibilities to maximize the expectation of the complete likelihood \ref{eq:complete loglikelihood} under these posterior probabilities
\begin{align}
\widehat{\Theta}^{(t+1)} & = \underset{\Theta}{\text{argmax }} E_{C,\mathbf{1}^{\prime}G|y,t,\widetilde{G},\widehat{\Theta}^{(t)}}\left[\sum_{i=1}^{N}\log\left[\tau_i(c,d;p,q)f\left(y_i; \theta_c, d\right) \right]\right]\\
& = \underset{\Theta}{\text{argmax }} \sum_{i=1}^N \sum_{c} \sum_{d=0}^{N-1} \widehat{\gamma}_{icd}^{(t+1)}\log\left[ \tau_{i}(c,d;p,q)f\left(y_i;\theta_c,d\right)\right].
\label{eq:EM Mstep}
\end{align}
For example, under the linear model \ref{eq:f linear}, we can compute closed form updates for the regression parameters:
\begin{align}
\widehat{\beta}_c^{(t+1)} & = \frac{\overline{dy}_{c} - \overline{d}_{c}\overline{y}_{c}}{\overline{d^2}_{c} - \overline{d}_{c}^2}\\
\widehat{\alpha}_c^{(t+1)} & = \overline{y}_{c} - \widehat{\beta}_c^{(t+1)}\overline{d}_{c}\\
\widehat{\sigma}^{2(t+1)} & = \frac{1}{N}\sum_{i=1}^{N}\sum_{c,d}\left[\widehat{\gamma}_{icd}^{(t+1)}\left(y - \widehat{\alpha}_c^{(t+1)} - \widehat{\beta}_c^{(t+1)}d\right)^2\right]
\label{eq:EM Mstep Normal Eqvar}
\end{align}
where $\overline{d}_{c} = \frac{\sum_{i=1}^N\widehat{\gamma}_{icd}^{(t+1)}d}{\sum_{i=1}^N\widehat{\gamma}_{icd}^{(t+1)}}$, $\overline{y}_{c} = \frac{\sum_{i=1}^N\widehat{\gamma}_{icd}^{(t+1)}y_i}{\sum_{i=1}^N\widehat{\gamma}_{icd}^{(t+1)}}$, $\overline{d^2}_{c} = \frac{\sum_{i=1}^N\widehat{\gamma}_{icd}^{(t+1)}d^2}{\sum_{i=1}^N\widehat{\gamma}_{icd}^{(t+1)}}$, and $\overline{dy}_{c} = \frac{\sum_{i=1}^N\widehat{\gamma}_{icd}^{(t+1)}dy_i}{\sum_{i=1}^N\widehat{\gamma}_{icd}^{(t+1)}}$. Note the similarly of our linear model estimators to those obtained from weighted least squares. Updates for the mismeasurement parameters $p$ and $q$ cannot be computed in closed form but can be solved for using a general optimizer such as \texttt{optim} in R. Estimates of the mean outcomes under each exposure condition (\ref{eq:mix estimate}) are functions of the model parameters and can be calculated accordingly.

Given the likelihood (\ref{eq:complete loglikelihood}) is bounded and satisfies mild smoothness conditions, as well as sufficiently many runs of the EM algorithm, we should be able to find the global optima and obtain the MLE $\widehat{\Theta}_{EM} = \widehat{\Theta}_{MLE}$ \citep{mclachlan1988mixture}. We can consider the consistency of $\widehat{\Theta}_{MLE}$ under the scenario we had access to comparable experiments on many networks, and that the outcomes for each experiment arise from the mixture model (\ref{eq:loglikelihood}) with the same set of parameters $\Theta^*$. The major condition from \citet{wald1949note} needed to ensure consistency of $\widehat{\Theta}_{MLE}$ is the identifiability of the mixture model on a non-zero probability set of the subjects in these experiments. In the case of our linear model for the potential outcomes (\ref{eq:f linear}), consistency requires regular variation in the observed in-degrees and exposure conditions. Building off of Proposition 1, a sufficient condition for the consistency of $\widehat{\Theta}_{MLE}$ would be to observe infinitely many subjects with at least two distinct observed in-degrees $\tilde{d}$ under each observed exposure condition.

While the EM algorithm does not provide standard errors and confidence intervals for our estimate $\widehat{\Theta}_{EM}$ and functions thereof (such as the mean outcome estimates), bootstrap methods have been used in the context of other mixture models to approximate these quantities \citep{basford1997standard}. In particular, we consider the parametric bootstrap, which consists of the following steps:
\begin{enumerate}
    \item Generate samples $\{y^{(1)},y^{(2)},..,y^{(m)}\}$ from the fitted model given by (\ref{eq:loglikelihood}) with parameters $\widehat{\Theta}_{EM}$, holding the treatment assignment vector $t$ and observed network $\widetilde{G}$ fixed.
    \item Estimate $m$ sets of parameters $\{\widehat{\Theta}_{EM}^{(1)},\widehat{\Theta}_{EM}^{(2)},..,\widehat{\Theta}_{EM}^{(m)}\}$ using the EM algorithm on the generated samples $\{y^{(1)},y^{(2)},..,y^{(m)}\}$.
    \item Calculate Monte-Carlo estimates of the standard errors and/or confidence intervals for $\widehat{\Theta}_{EM}$ using the parameters estimates $\{\widehat{\Theta}_{EM}^{(1)},\widehat{\Theta}_{EM}^{(2)},..,\widehat{\Theta}_{EM}^{(m)}\}$.
\end{enumerate}

\subsection{Simulation Study} \label{section:Simulations}

In this section, we apply our mixture model to simulated experiments run over the households of 75 Indian villages, collected by and described in detail in \citep{banerjee2013diffusion}. Within each village, we consider an experiment with two treatment levels (``treatment" and ``control") and treatment interference on the outcome of interest through the household network of borrowing and lending money.  \footnote{Data are available from \url{https://dataverse.harvard.edu/dataset.xhtml?persistentId=hdl:1902.1/21538}.}

The 75 villages range in size from 77 to 356 households, with an average of slightly under 200. We take the true spillover network in each village to be the reported network of borrowing and lending, with a link between two households if there is any monetary borrowing or lending between the two. While these networks may themselves be mismeasurements of the actual borrowing/lending networks for each village, they are nonetheless a useful laboratory in which to explore the performance of our proposed estimator.  In particular, they exhibit realistic properties one may expect in networks, such as small world phenomenon, significant clustering, and substantial variation in degrees. When excluding households with zero degree, the average number of households in a village is about 170.

We run 10 simulated experiments on each village, each of which consists of the following steps:
\begin{enumerate}
    \item Independently assign each household to treatment with probability $0.25$.
    \item Calculate the exposure condition $C_i$ for each household $i$ using the true network $G$.
    \item Generate outcome of interest $y_i$ for each household according to (\ref{eq:f linear}) with parameters $\alpha = (0,0.25,0.5,1)$, $\beta = (0.05,0.1,0.05,0.1)$, and $\sigma^2=0.25$.
    \item Observe a mismeasured network $\widetilde{G}$, where each edge in the true network is observed independently with probability $1-p$, while false edges in the network are observed independently with probability $qd$, where $d$ is the density of true graph $G$\footnote{This specification for the mismeasurement parameter governing the observance of false edges is easier to compare across networks with varying densities. $q=0.5$ implies observing a number of false edges equal to about half the actual number of true edges.}. Simulations are repeated for every pair of mismeasurement probabilities $(p,q)$ with $p\in\{0,\frac{1}{8},\frac{1}{4},\frac{3}{8},\frac{1}{2}\}$ and $q\in\{0,\frac{1}{8},\frac{1}{4},\frac{3}{8},\frac{1}{2}\}$.
    
\end{enumerate}
Across the 10 simulations and 75 villages, we average 45 households with no exposure to treatment (under the true network), 82 households with indirect exposure to treatment, 15 households with direct exposure to treatment, and 28 households with full exposure to treatment. The challenge is to derive estimates of the mean outcome under each exposure condition (\ref{eq:mu}) despite observing a mismeasured network $\widetilde{G}$ and thus mismeasured exposure conditions. Some of our exposure conditions can be quite challenging; for example, when $p=q=0.5$, on average half of the true edges are not observed, but instead are ``replaced" with roughly the same number of false edges.

Estimation proceeds as described in Section \ref{section:Estimation}. We choose the parameters of the beta-binomial distribution over the true degrees (recall the discussion following equations \ref{eq:num trt friends} and \ref{eq:num non-trt friends}) by taking $\mu$ to be the density of the true network and choosing dispersion parameter $\rho$ such that the second moment of the beta-binomial distributions matches that of the the observed degree distribution. This simulation scenario is consistent with a setting where we have \textit{a priori} expectations on the density of the true network of spillovers.  In Figure \ref{fig:Mean Estimates No Direct 2} located in Appendix \ref{section:Figures}, we present results where $\mu$ and $\rho$ are chosen solely by matching the first two moments of the observed degree distribution. Our results are slightly worse under the purely empirical specification but still represent a substantial improvement over the Horvitz-Thompson approach that does not account for any mismeasurement.

\begin{figure}[p]
\centering
\includegraphics[scale=0.65]{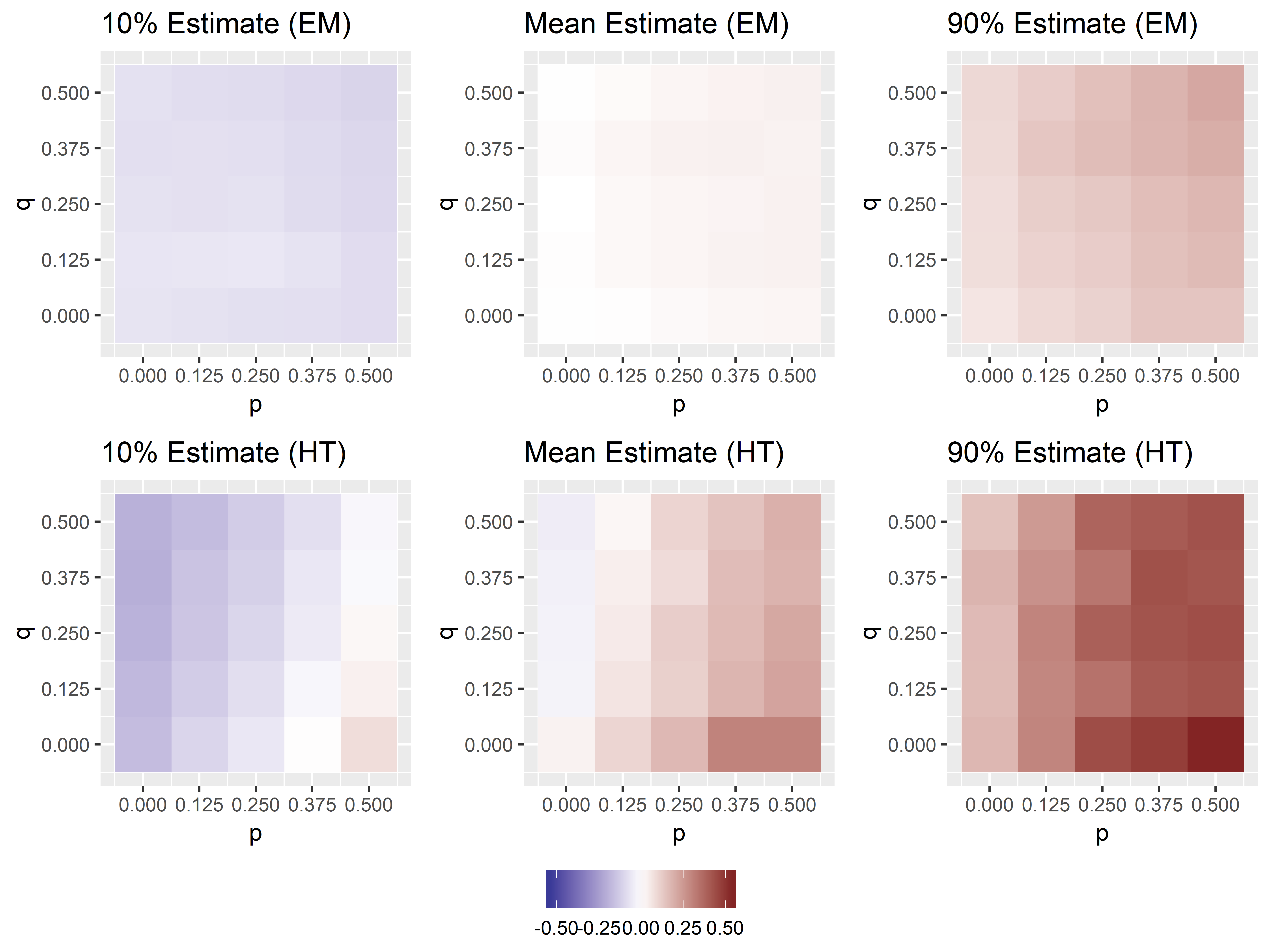} 
\includegraphics[scale=0.65]{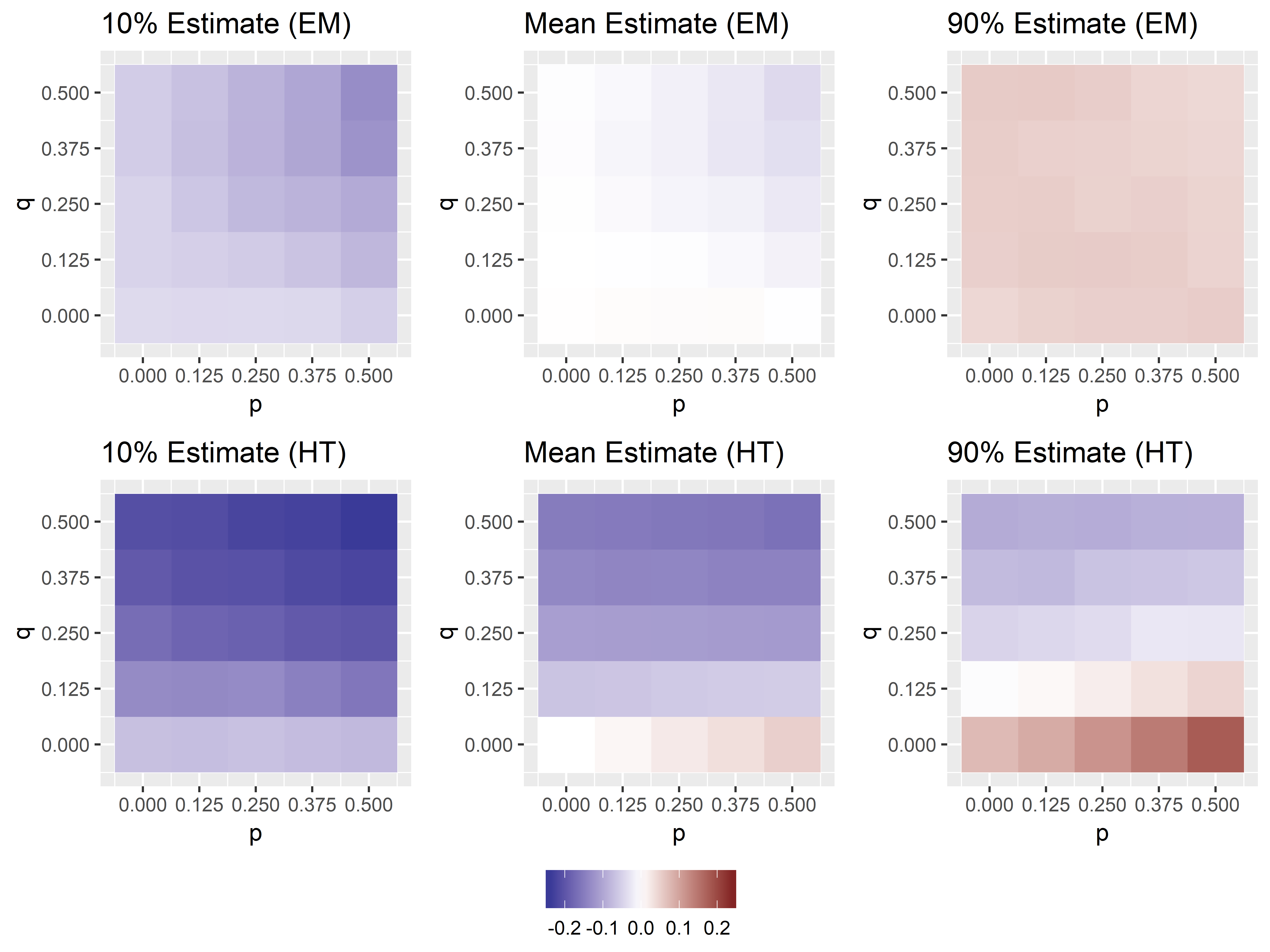} 
\caption{\small Estimates of the mean outcome of the no exposure (top) and indirect exposure (bottom) conditions from their true values under varying mismeasurement levels ($p$,$q$) for the network. Estimates obtained from our model using the EM algorithm are compared against estimates from the Horvitz-Thompson estimators assuming no mismeasurement in the network. The 0.1 and 0.9 quantiles are provided for both methods to give a sense of the variability in these estimates. Note the color gradient scales are different for the two exposure conditions.}
\label{fig:Mean Estimates No Direct}
\end{figure}

\begin{figure}[p]
\centering
\includegraphics[scale=0.65]{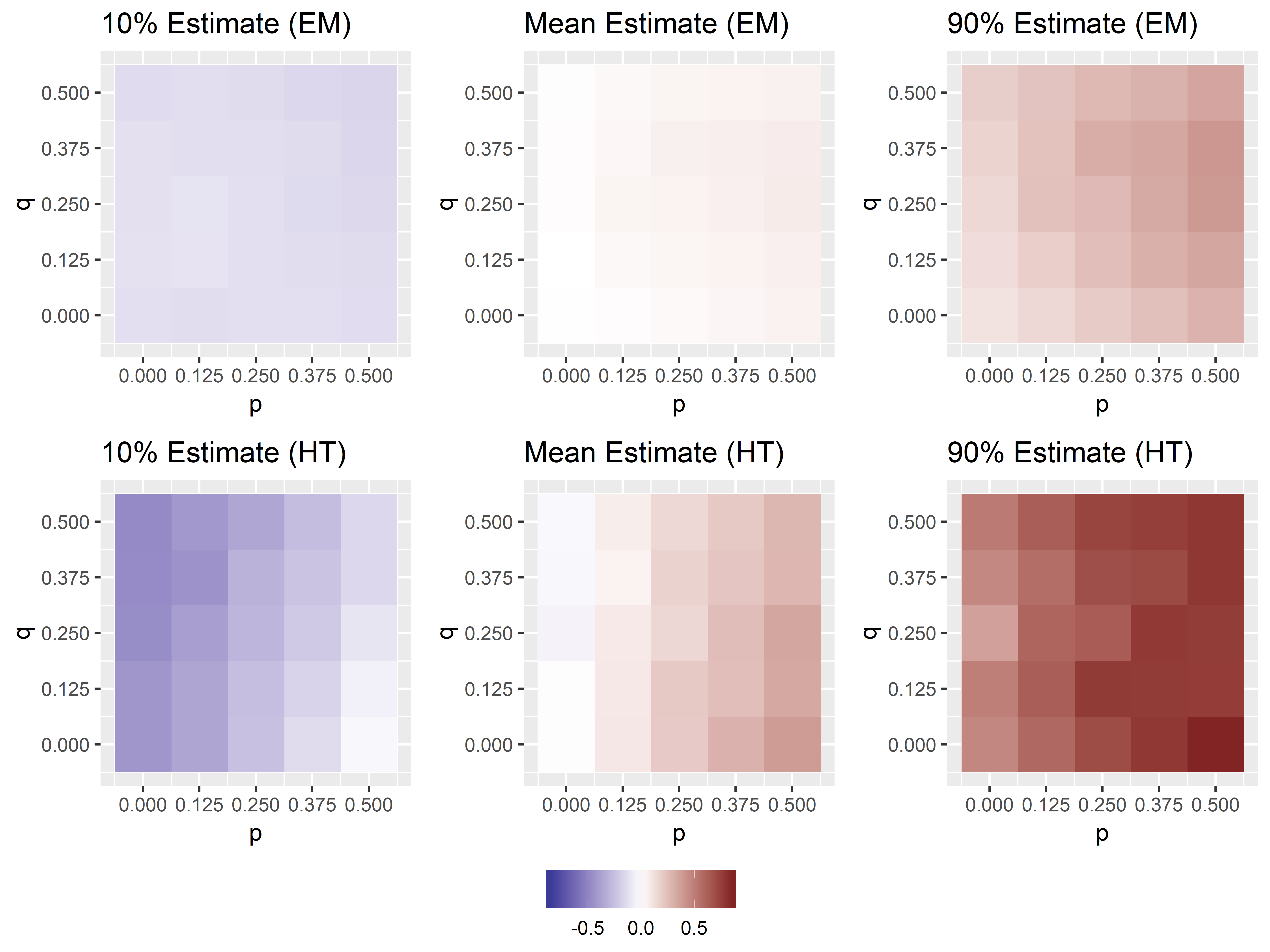} 
\includegraphics[scale=0.65]{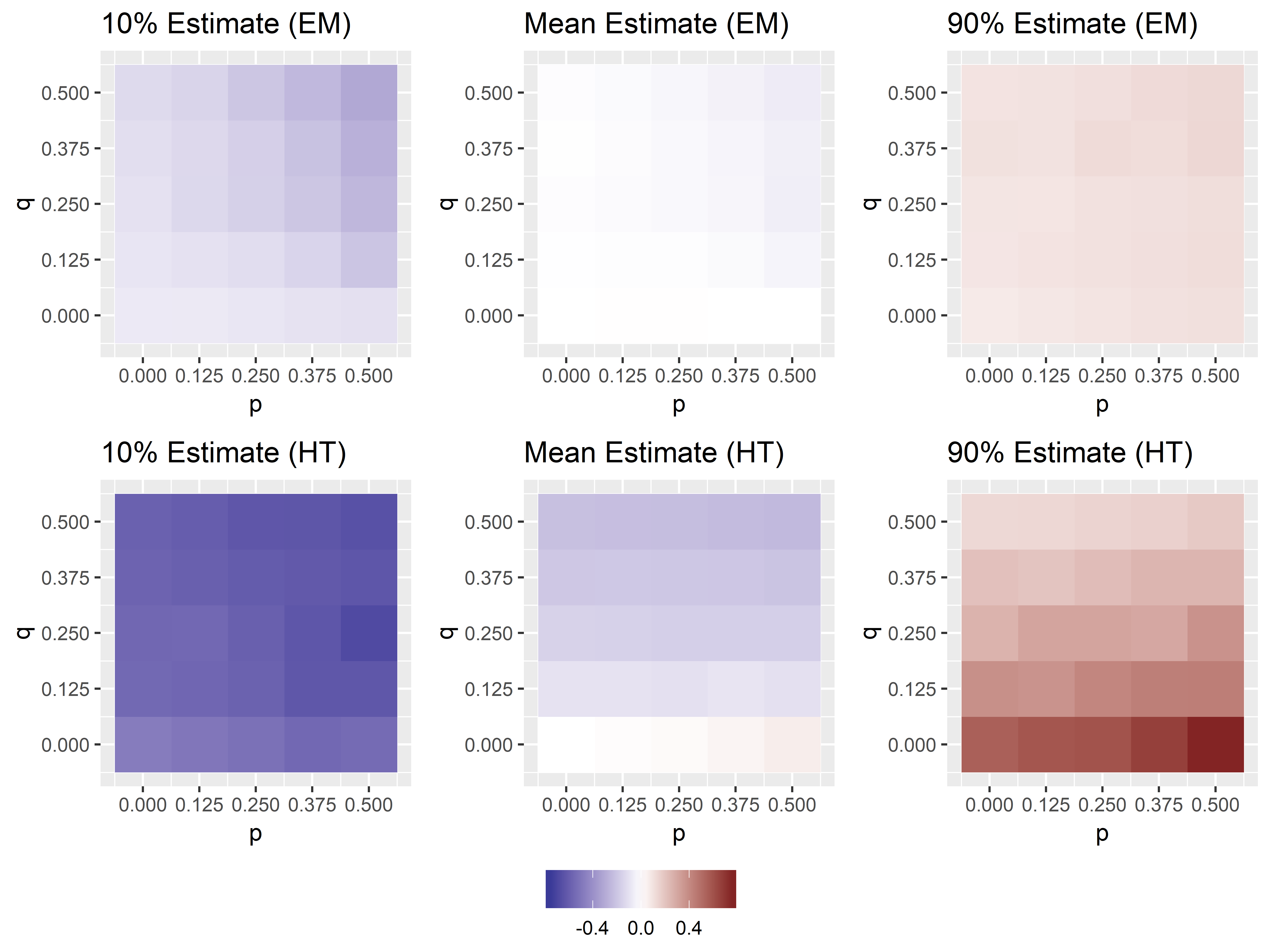} 
\caption{\small Estimates of the deviation of the mean outcome of the direct exposure (top) and full exposure (bottom) conditions from their true values under varying mismeasurement levels ($p$,$q$) for the network. Estimates obtained from our model using the EM algorithm are compared against estimates from the Horvitz-Thompson estimators assuming no mismeasurement in the network. The 0.1 and 0.9 quantiles are provided for both methods to give a sense of the variability in these estimates. Note the color gradient scales are different for the two exposure conditions.}
\label{fig:Mean Estimates Direct}
\end{figure}

In Figures \ref{fig:Mean Estimates No Direct} and \ref{fig:Mean Estimates Direct}, we present heat maps to compare the estimates obtained by the EM algorithm to the Horvitz-Thompson estimates that do not take into account missingness in the network, at varying levels of unreported true links ($p$) and falsely observed links ($q$). Positive deviations from the true mean outcomes are denoted in red while negative deviations are denoted in blue, with the intensity of the colors corresponding to the size of the deviation.  

We first comment on bias in the Horvitz-Thompson estimates, corresponding to the analytical bias formula described in section \ref{section:AS model mismeasurement}.  To begin, we consider the bias in estimating indirect treatment effects, i.e., the mean outcome among individuals who are not treated themselves, but have at least one treated network member.  (Given our assumption that an individual's own treatment status is observed without error, the logic for the full treatment condition -- individuals who are treated themselves and know a treated member -- is identical.)

First, consider the case where $p$ is zero, but $q$ is positive: all true links are observed, but some
false links are as well.  So some people that appear to be indirectly treated are not actually indirectly treated, and the observed indirect treatment effect will be biased downward as a result.  This bias increases with $q$, as the fraction of people observed as treated who are actually treated falls.

Now, alternatively, consider the case where $q$ is zero, but $p$ is positive: all observed links are true links, but some true links are dropped.  In this case, if someone is observed to be indirectly treated, they actually are indirectly treated.  However, there is still bias in the Horvitz-Thompson estimator, since low-degree individuals who are actually indirectly treated are particularly likely to be mis-classified as not indirectly treated.  Under the assumption that degree is positively correlated with the outcome, disproportionately dropping low degree individuals from the indirectly treatment will bias the estimated treatment effect upwards.

We turn now to estimates of the zero exposure treatment (individuals who are not treated themselves, and do not have a treated network member either).  The logic is similar.  If we first fix $q$ at zero and increase $p$ -- thereby dropping true links -- then some people that we think are \emph{not} indirectly treated are actually indirectly treated.  Since indirect treatment has a positive effect, this mismeasurement biases the estimated mean of no treatment upwards.

If we instead add false links by fixing $p$ at zero and increasing $q$, then every individual that we observe  as not treated is actually not treated.  However, the individuals who are observed as not treated even after we add false links are disproportionately low degree.  Given that degree is positively correlated with the outcome, adding these low degree individuals biases the estimated mean in the no treatment group downwards.

By contrast, the estimates of mean outcomes given by the EM algorithm are quite reasonable across the varying levels of mismeasurement $p$ and $q$ considered, and represent a substantial improvement over the Horvitz-Thompson estimates. Differences in performance across the various mismeasurement levels are much more muted, at least across the different levels of mismeasurement considered in our simulations.

\begin{figure}[h]
\centering
\includegraphics[scale=0.65]{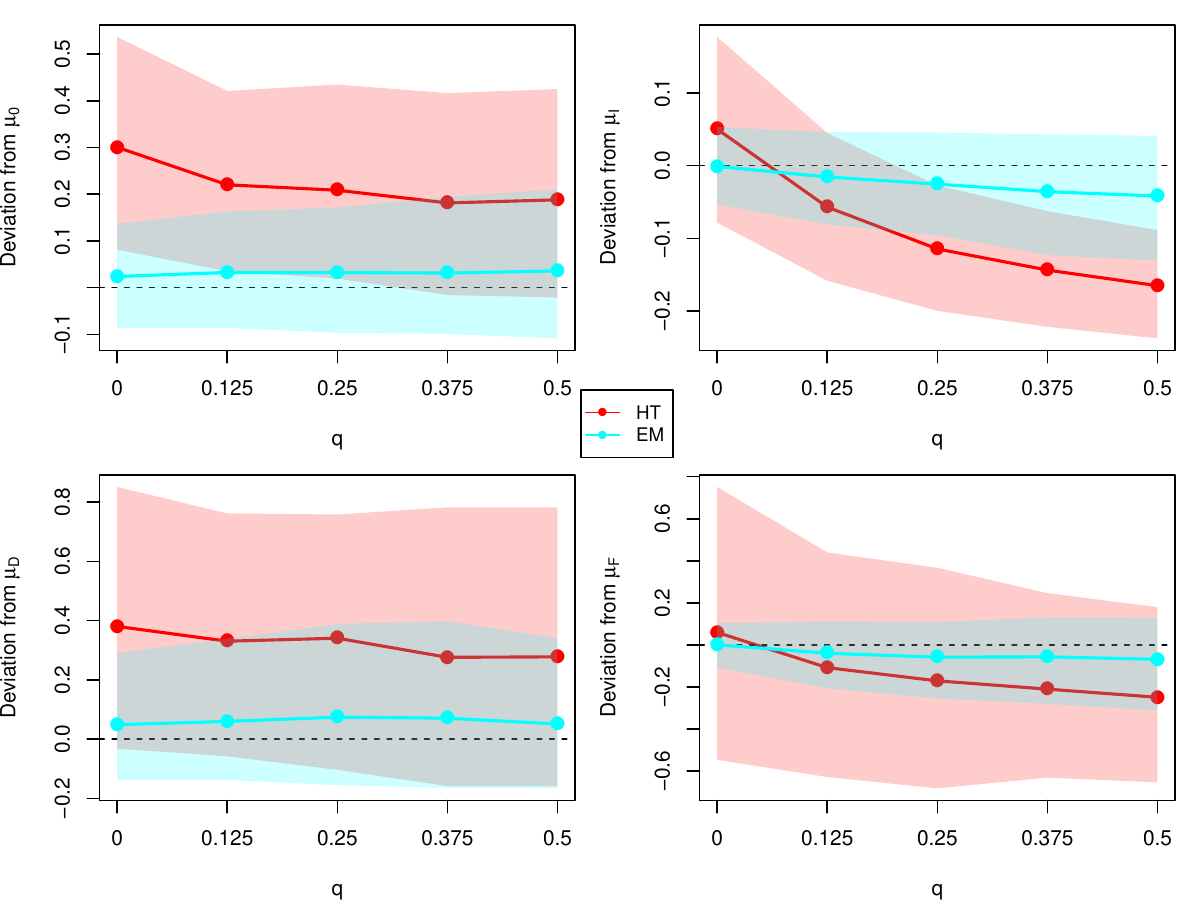} 
\caption{\small Estimates of the deviation of the mean outcome of each exposure conditions (top left: no exposure, top right: indirect exposure, bottom left: direct exposure, bottom right: full exposure) from their true values under $p=0.5$ and varying $q$ from 0 to 0.5. Estimates obtained from our model using the EM algorithm are compared against estimates from the Horvitz-Thompson estimators assuming no mismeasurement in the network. The 0.1 and 0.9 quantiles are shaded for both methods to give a sense of the variability in these estimates.}
\label{fig:Mean Estimates p=0.5}
\end{figure}

\begin{figure}[h]
\centering
\includegraphics[scale=0.65]{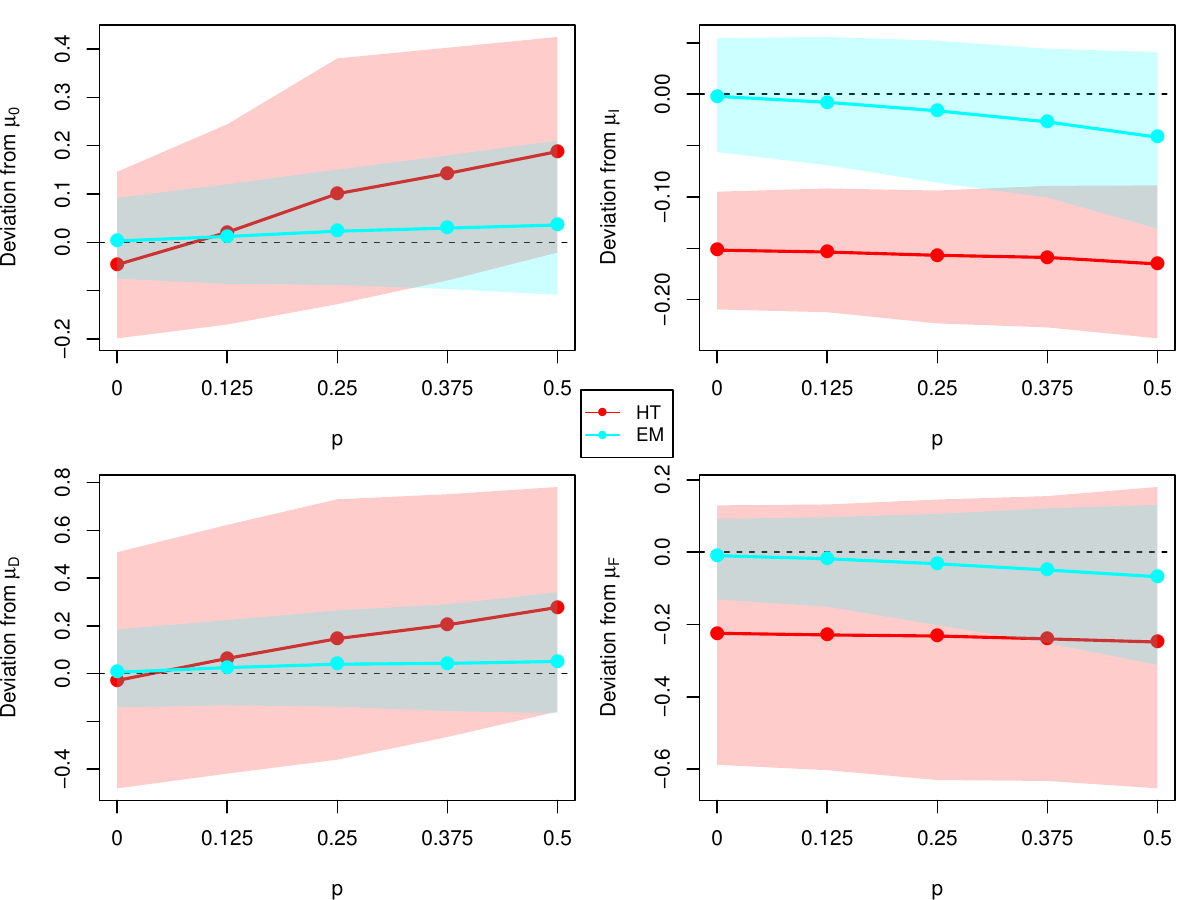} 
\caption{\small Estimates of the deviation of the mean outcome of each exposure conditions (top left: no exposure, top right: indirect exposure, bottom left: direct exposure, bottom right: full exposure) from their true values under $q=0.5$ and varying $p$ from 0 to 0.5. Estimates obtained from our model using the EM algorithm are compared against estimates from the Horvitz-Thompson estimators assuming no mismeasurement in the network. The 0.1 and 0.9 quantiles are shaded for both methods to give a sense of the variability in these estimates.}
\label{fig:Mean Estimates q=0.5}
\end{figure}

To examine these results in more detail, we focus on the two cases presented in Figures \ref{fig:Mean Estimates p=0.5} and \ref{fig:Mean Estimates q=0.5}, where we fix $p=0.5$ and vary $q$ and fix $q=0.5$ and vary $p$ respectively. Estimates from our method are presented in cyan while estimates from the Horvitz-Thompson are presented in red. In general, estimates from our method exhibit considerably less bias and are simultaneously have less variance. We find that our method has slightly higher levels of bias and variance for higher levels of mismeasurement, which is consistent with the idea that for higher $p$ and $q$ there is larger uncertainty over the true network (\ref{eq:mixture probabilities}) and thus our results are more dependent on the assumed beta-binomial model over the true degree distribution. Even if $\mu$ and $\rho$ are chosen to match the true degree distribution, the beta-binomial model still represents a (higher-order) deviation from the true degree distribution for real-life networks. The direction of the bias, upwards for $\widehat{\mu}_{0}$ and $\widehat{\mu}_{D}$ towards $\mu_{I}$ and $\mu_{F}$ respectively and the reverse for $\widehat{\mu}_{I}$ and $\widehat{\mu}_{F}$, are a product of imperfectly learning the latent exposure conditions, especially in these higher uncertainty settings.

\section{Diffusion of insurance information between farmers}
\label{section:Application}

\citet{cai2015social} study the adoption decisions of rice farmers in rural China in regards to weather insurance. Typically  the take-up rates for insurance are low even amongst these farmers in the presence of heavy government subsidies. \citet{cai2015social} examined how difficulties communicating the benefits of the product could be modulated if information about insurance comes via a farmer's peers. In conjunction with the introduction of a new weather insurance product, researchers randomized about 5000 households across 185 rural villages into two rounds of information sessions about the new insurance product. Sessions were held in two rounds three days apart, and could either be ``simple" sessions just describing the product or longer ``intensive" sessions which also emphasized the expected benefits from insurance. Drop out was not a major issue in this experiment, with an overall session attendance rate of about 90\%.

One specific question the authors were interested in was how insurance take-up and knowledge for households assigned to second round sessions were affected by whether or not they had friends assigned to first round intensive sessions. ~\cite{cai2015social} construct three different measures of social connectivity.  First, before the experiment, each household was asked to list five friends whom they most frequently discussed production or financial issues with.
In general, prompting respondents to list five friends can censor the number of connections for individuals with high in-degree, as well as cause the reported network to contain some weaker connections that would otherwise be unreported. However, this concern may be relatively mild in this case, as the authors conducted a pilot study in two villages where the number of friends was uncensored and found 96\% of farmers reported either four or five connections. Most of the paper's results use this reported network, which, borrowing their language, we will term the ``general network measure."  Second, ~\cite{cai2015social} define a ``strong" network measure where non-reciprocal connections are dropped.  That is, two individuals are connected only if each person lists the other among their five friends.  The third measure is a ``weak" network measure which adds second-order connections (``friends of friends") to the general network measure. Both specifications differ quite drastically from the general network measure and are indicative of low rates of reciprocity and transitivity in the reported network; farmers average a single connection under the strong network measure (with a mode of 0) and 16 connections under the weak network measure.

To measure insurance knowledge, each household completed a five question test after the experiment and were scored from 0-5. For households assigned to second round sessions, ~\cite{cai2015social} found that having a friend who was assigned a first round intensive session had the same (statistically significant) benefit for insurance knowledge, as measured by score on the five question test, as personally being assigned an intensive session in the second round. 
 
 We model each household's score on the insurance test as arising from a binomial with five independent questions and a probability of getting a question right depending on the household's treatment exposure condition as well as their degree in the network.  
\begin{equation}
    \text{score} \sim \text{Bin}(5,\text{expit}(\alpha_c + \beta_c d))
    \label{eq: insurance score}
\end{equation}
 In the context of this experiment, being directly treated corresponds to a farmer being invited to the intensive training.  Being indirectly treated corresponds to having a network member invited to the intensive training.  To produce comparable estimates to the linear specification (2) presented in Table 5 of~\cite{cai2015social}, we estimate the mean outcome under each exposure condition and calculate various contrasts using these means. The effect of personally being invited to a intensive session can be calculated as $\widehat{\mu}_{D} - \widehat{\mu}_{0}$, the effect of having a friend invited to a first round intensive session (which we denote as  ``Network Intensive") is calculated as $\widehat{\mu}_{I} - \widehat{\mu}_{0}$, and the interaction of these effects is given by $\widehat{\mu}_{0} + \widehat{\mu}_{F} - \widehat{\mu}_{I} - \widehat{\mu}_{D}$.

 We focus on the effects of having a network member attend the first round intensive training, the network intensive condition.  Tables with estimates for all of the conditions are in Appendix~\ref{section:tables}. Figure~\ref{fig:betacoef} shows the estimates and uncertainty intervals for being in the network intensive condition.  A positive value indicates that having a person in your network receive the intensive first round treatment increases your knowledge about the insurance product.  Each of the three plots represents a different estimation strategy (assuming no mismeasurement, our proposed EM method, and our method utilizing covariates) for the three possible measures of network connections. Thick lines represent one standard error and thin lines represent the width of two standard errors.
 
 The leftmost panel of Figure~\ref{fig:betacoef} gives the estimates without adjusting for any potential errors in mismeasurement in the graph.  The impact of being in the network intensive condition on insurance knowledge is positive and more than two standard deviations from zero for both the general and strong measures, though the strong estimate is noticeably larger than the general measure.  For the weak measure, the estimate is now no longer two standard errors from zero.  This observation is striking since it indicates that an investigator who defines the graph using the weak measure would come to a substantively different conclusion than one using the other two measures.
 
 Moving now to the middle panel, which shows the results from our proposed approach\footnote{The beta-binomial distribution over the true degrees is initialized with the same mean as the observed network (reflecting the results from the pilot study) and overdispersion parameter 0.0005. This overdispersion parameter was chosen based on examining the variation of the degrees in the Indian village data used in the simulations. Note the observed degree-distribution in farmers' network exhibits considerably less variation than even a binomial distribution, and thus is not particularly informative for choosing our prior distribution.}, the alignment between the three network measures is much more consistent. As described above, we expect that the rate of censoring and over-reporting is relatively low since individuals listed approximately the same number of contacts in pilot studies when the given a limit of five.  We would expect, therefore, that our method will nearly replicate the results from~\cite{cai2015social}.  %Specifically,
\begin{figure}
    \centering
    \includegraphics[width=.3\textwidth]{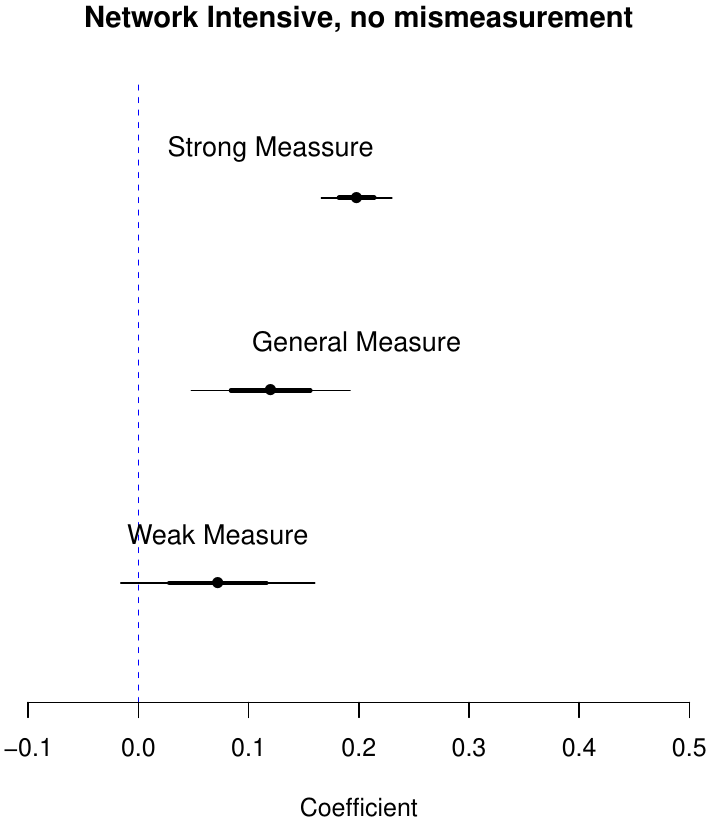}
    \hspace{10pt}
    \includegraphics[width=.3\textwidth]{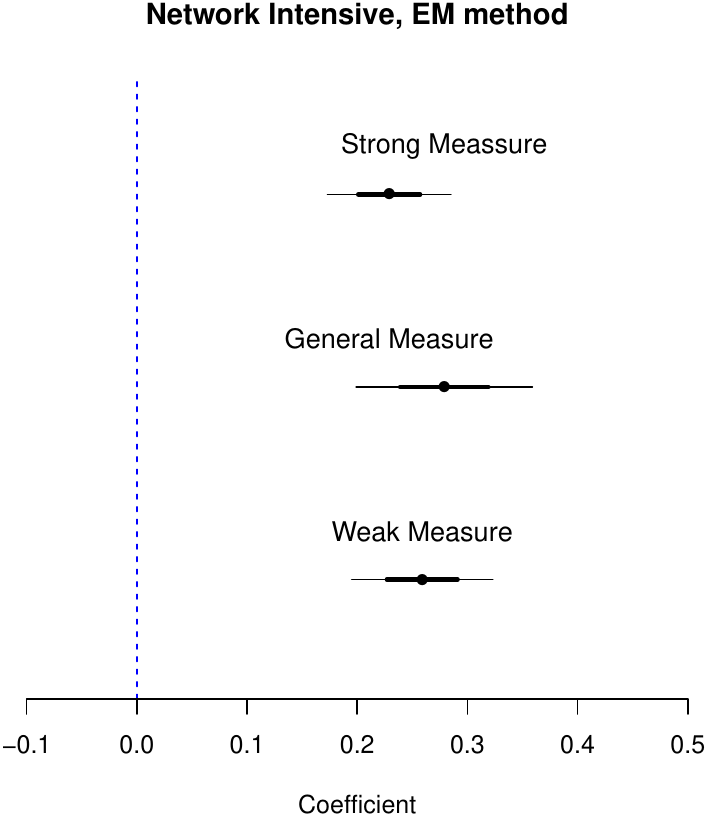}
    \hspace{10pt}
    \includegraphics[width=.3\textwidth]{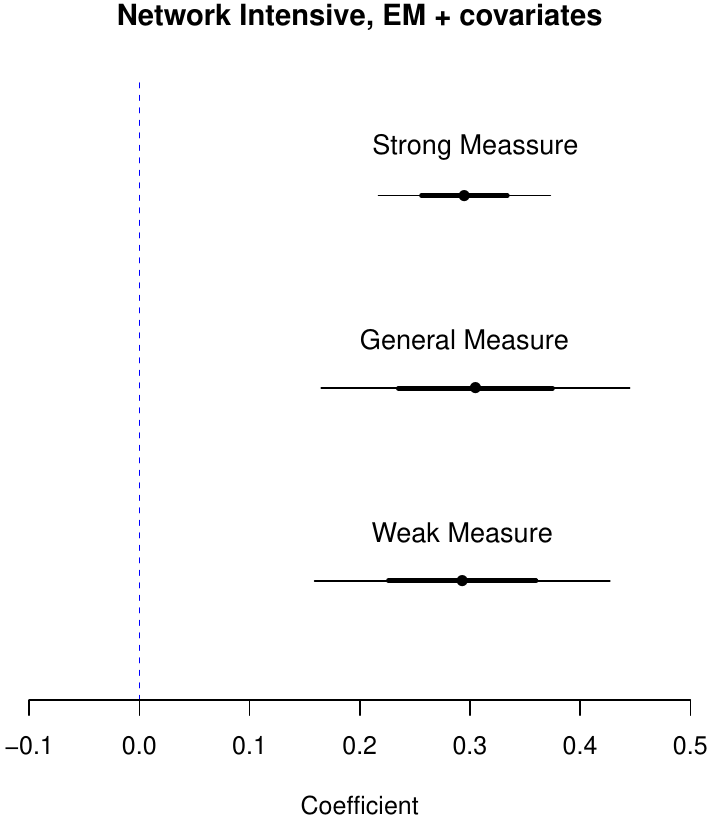}
    \caption{Differences in insurance knowledge for farmers assigned to second round sessions based on whether they had a friend attend a first round intensive session. We compare our method to results if we assume there is no mismeasurement in the network for three network measures.}
    \label{fig:betacoef}
\end{figure}
Finally, we consider an extension of our model (\ref{eq:mixture probabilities}) that allows for different levels of mismeasurement in the connections between farmers depending on whether or not they reside in the same village. About 99.4\% of reported connections are between farmers in the same village\footnote{There is substantial variation in the size of each village, which is entirely not reflected under the network measures considered}, while the remaining 0.6\%, so separately modeling the true degree within-village and out-of-village may lead to more accurate results. We introduce distinct parameters for in-village degree ($\mu_{in}$ and $\rho_{in}$) and out-of-village degree ($\mu_{out}$ and $\rho_{out}$), along with respective mismeasurement parameters $p_{in}$, $q_{in}$, $p_{out}$, and $q_{out}$. Note that there is a potential variance trade-off when introducing additional parameters to our model, so sample size concerns must also be considered. Estimation proceeds as described in Section \ref{section:Estimation}, with the additional complication that the general purpose optimizer must maximize over all four mismeasurement parameters at once. Estimates from this extension are largely similar to those obtained from our method ignoring the difference between in-village and out-of-village ties, perhaps due to the lack of out-of-village ties. However, the corresponding standard errors are substantially larger since we have introduced additional parameters to estimate.

\begin{figure}
    \centering
    \includegraphics[width=\textwidth]{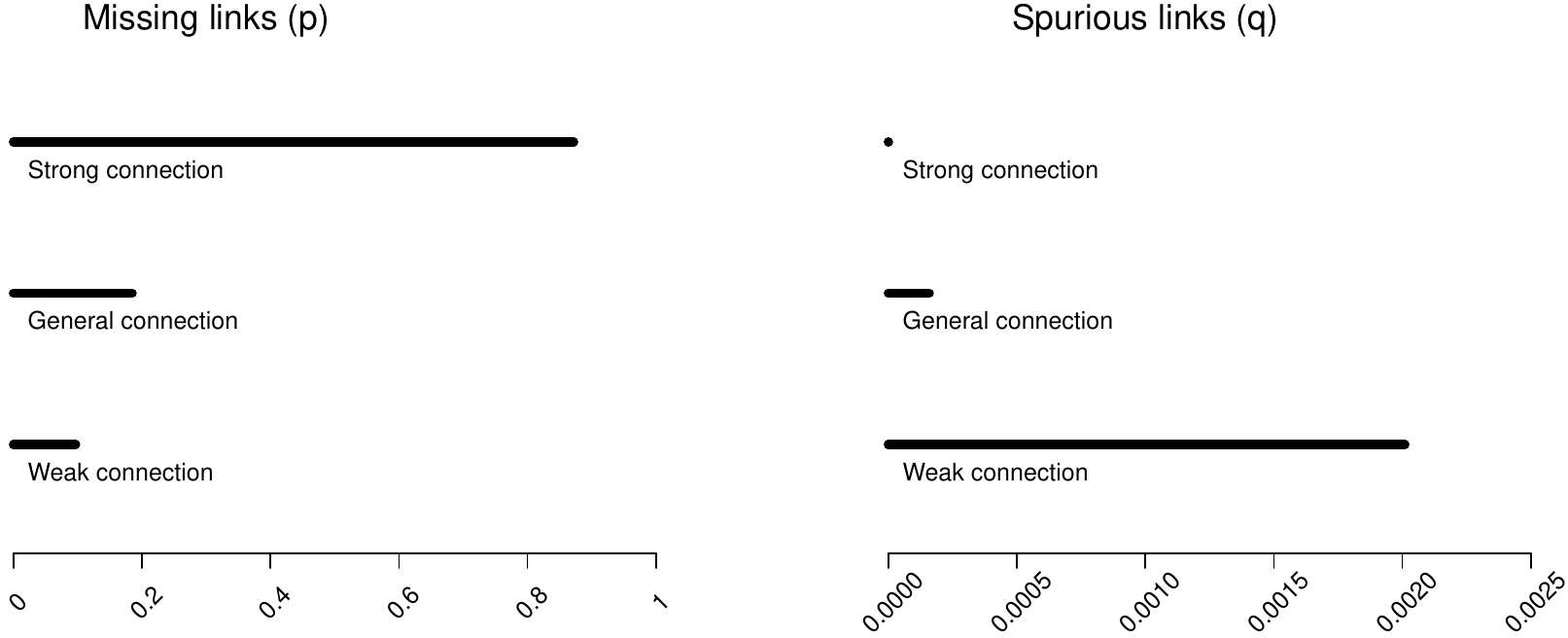}
    \caption{Point estimates of mismeasurement.  The left panel gives the estimated fraction of missing links ($p$) for each of the three network definitions.  The right panel givens the same information for spurious links ($q$).}
    \label{fig:pqpic}
\end{figure}

To further illustrate the functioning of our method, we display the fraction of missing and spurious links, as estimated by our approach, in the three network specifications.  Figure~\ref{fig:pqpic} plots the point estimate of the fraction of missing links ($p$) and fraction of spurious links ($q$) for each network definition.  As expected, the fraction of missing links is substantially higher with the most stringent definition of a tie than in the weakest definition of a connection.  The opposite pattern appears in the right side panel for the fraction of links that are spurious.\footnote{Note that the estimated $q$'s are an order of magnitude lower than the estimated $p$'s.  This is to be expected, as $q$ gives the probability of a given hypothetical link falsely appearing.  So unless the size of the network is large relative to the total population, $q$ being anything other than very small would mean that a huge fraction of observed links are actually false.  And, as previously discussed, networks are allowed to extend between villages in the \citet{cai2015social} data, so the universe of potential connections is quite large.}  In both cases, the general measure is in between the two extremes. 

\section{Discussion} \label{section:Discussion}

Experimental inference on social networks presents distinct challenges; not only are subjects' outcomes affected by the treatment assignments of other subjects, but this treatment interference is often of direct interest. Existing methodology for estimating treatment effects in this setting requires a precise measurement of the network of interest, which can be a difficult assumption given the many decisions inherent in the data gathering process as well as imposing a large financial burden. In this paper, we present a class of mixture models that can accurately estimate treatment effects when the network of interest is not accurately measured, assuming that the noise in the network is (conditionally) random and relying on additional assumptions about the parametric form for the treatment exposure conditions and the density of the true, latent network.

\begin{spacing}{1}
\bibliographystyle{abbrevnamed}
\bibliography{bibliographies/NetworkSpillovers}
\end{spacing}

\appendix

\section{Additional Figures} \label{section:Figures}

\begin{figure}[p]
\centering
\includegraphics[scale=0.65]{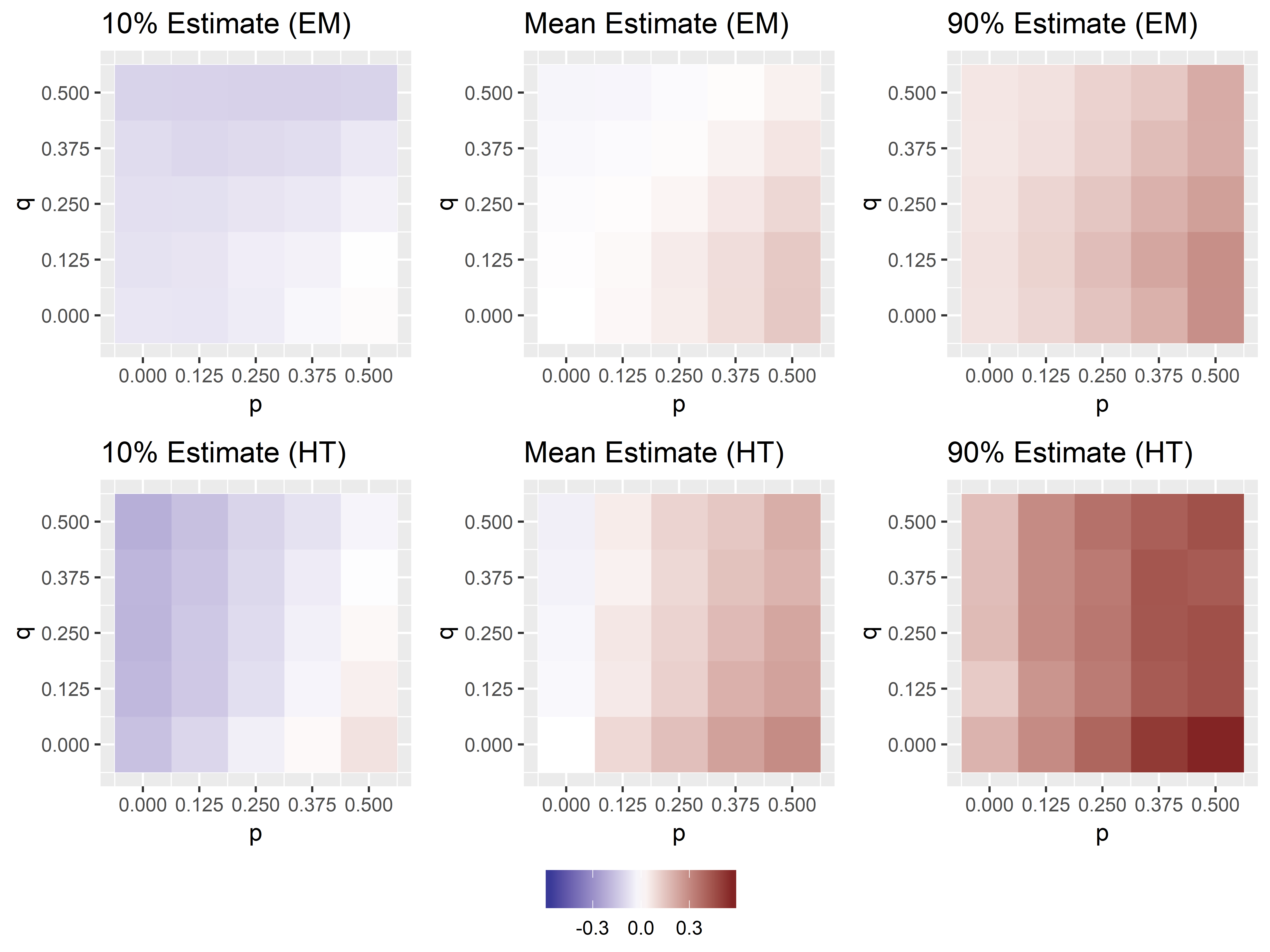} 
\includegraphics[scale=0.65]{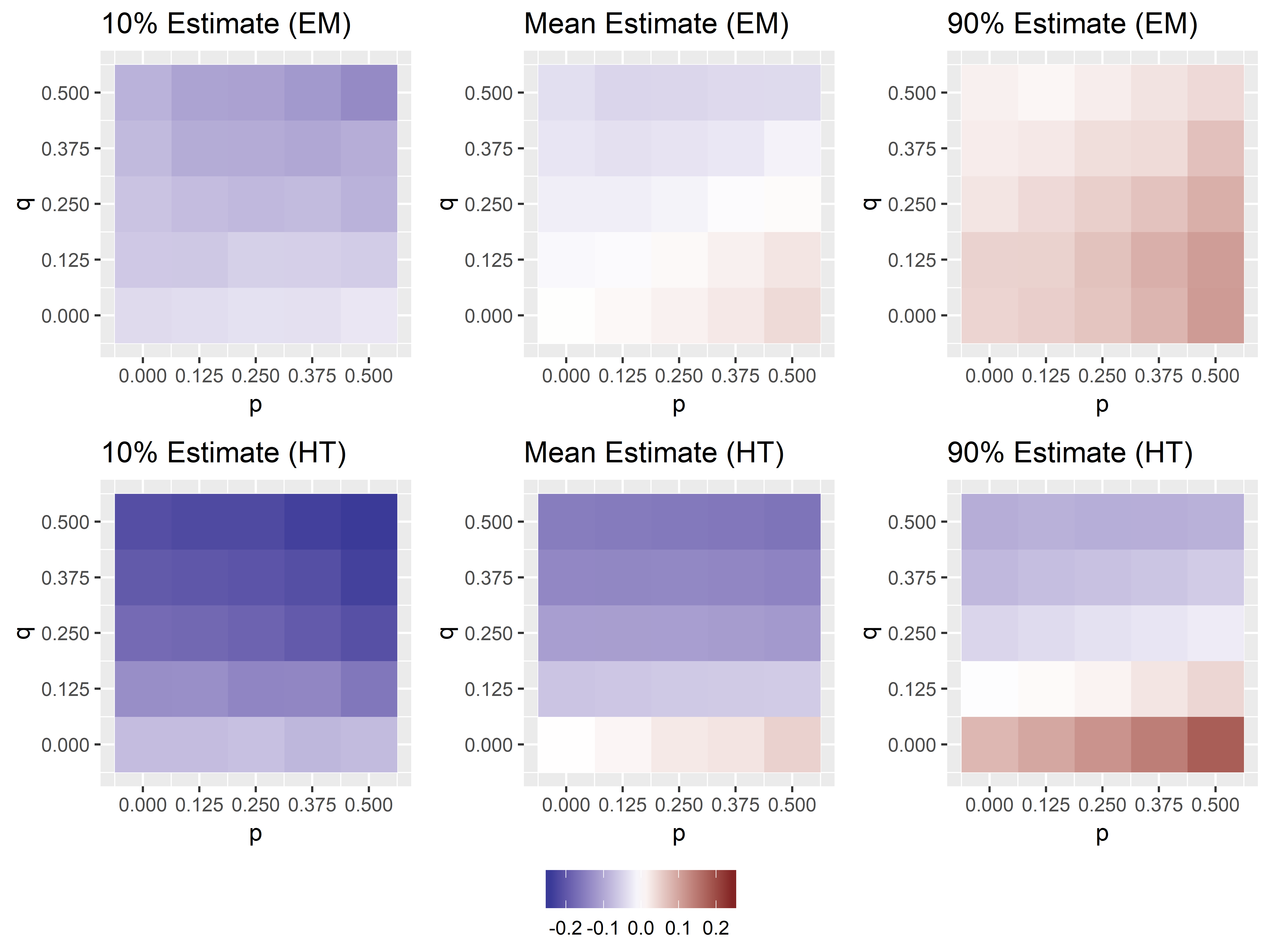} 
\caption{\small Estimates of the mean outcome of the no exposure (top) and indirect exposure (bottom) conditions from their true values under varying mismeasurement levels ($p$,$q$) for the network. Estimates obtained from our model using the EM algorithm are compared against estimates from the Horvitz-Thompson estimators assuming no mismeasurement in the network. The 0.1 and 0.9 quantiles are provided for both methods to give a sense of the variability in these estimates. Note the color gradient scales are different for the two exposure conditions.}
\label{fig:Mean Estimates No Direct 2}
\end{figure}

\pagebreak

\section{Proof of Identification} \label{section:IDproof}
It suffices to show identifiability of $\{\theta_{00},\theta_{01}\} = \{\theta_{00}^\prime,\theta_{01}^\prime\}$, since we assume direct treatment status can always be accurately ascertained. The exposure conditions $\{c_{00},c_{01}\}$ are only mismeasured with one another, as are $\{c_{10},c_{11}\}$.

Let us begin with the most general case, when both $\{p,q\}\in(0,1)$. In this situation, the probabilities $\tau_i$ are positive over all feasible true exposure conditions and degrees, regardless of the pair of observed degrees ($\tilde{d}_{t}$,$\tilde{d}_{nt}$). The only restriction on the support of these probabilities are that, under no indirect treatment, degree cannot be larger than $N-1-1^\prime t + t_i\equiv N_{nt,i}$ (otherwise there would have to exist a connection to a treated subject), and degree must be at least one for an individual to be indirectly treated. Mathematically, $\tau_i\left(c_{00},d;p,q\right) > 0$ for any $d$ satisfying $d\leq N_{nt,i}$ and $\tau_i\left(c_{01},d;p,q\right) > 0$ for any $d\geq1$.

At the other extreme, when there is no mismeasurement ($p=0$ and $q=0$), then the true exposure condition and degree match their observed counterparts. Mathematically, $\tau_i\left(c,d;p,q\right) > 0$ only for $d=\tilde{d}_{t} + \tilde{d}_{nt}$ and either $c=c_{00}$ if $\tilde{d}_{t} = 0$ or $c=c_{01}$ if $\tilde{d}_{t} > 0$. When exactly one kind of mismeasurement exists, the support of $\tau_i$ is limited, but to a lesser extent that when neither types of mismeasurement exist. When $p>0$ but $q=0$, true edges can be dropped but all observed edges also exist in the true network. Namely, any observed connection to a treated subject must exist in the true graph. For subjects with at least one of these connections $\tilde{d}_{t}>0$, the support of $\tau_i$ is limited to $c = c_{01}$ and $d\geq\tilde{d}_{t} + \tilde{d}_{nt}$. If instead we have $\tilde{d}_{t}=0$, $\tau_i$ is positive for $\tilde{d}_{nt} \leq d \leq N_{nt,i}$ when $c=c_{00}$ and $d\geq \tilde{d}_{nt} + 1$ when $c=c_{01}$. Lastly, when $q>0$ and $p=0$, the observed connections is a superset of the links in the true graph. Thus, when we observe no connections to treated subjects $\tilde{d}_{t}=0$, $\tau_i$ is only positive for $c=c_{00}$ and $d\leq\tilde{d}_{nt}$. When such an connection is observed, $\tau_i$ is positive for $c=c_{00}$ and $d\leq\tilde{d}_{t} + \tilde{d}_{nt}-1$ or $c=c_{01}$ and $1\leq d\leq\tilde{d}_{t} + \tilde{d}_{nt}$.

\begin{case}
$p > 0$, $q > 0$, and $\beta_{10} \neq 0$
\end{case}

For any pair of $\left(\tilde{d}_t,\tilde{d}_{nt}\right)$, the LHS is a mixture of normal distributions that includes $N-1$ distinct components with means $\alpha_{01}+\beta_{01}d$ and variance $\sigma^2$ for any $d$ from $\{1,...,N-1\}$. There are at most $N_{nt,i}+1 < N-1$ other mixture components corresponding to the $c_{00}$ terms. Following the generic identifiability of finite normal mixtures, the same component normals must exist on the RHS, with the same weights. For there to be at least $N-1$ distinct components on the RHS for both $\tilde{d}_t=0$ and $\tilde{d}_t>0$, we must have $p^\prime > 0$ and $q^\prime > 0$. On the LHS, we have $N-1$ components which are evenly spaced $\lvert\beta_{01}\rvert$ apart, while on the RHS we have $N-1$ components evenly spaced $\lvert\beta^\prime_{01}\rvert$ apart. Since there are fewer than $N-1$ other components on either side, these $N-1$ components must match, with $\lvert\beta_{01}\rvert = \lvert\beta^\prime_{01}\rvert$. This leads to two possibilities: we must have either $\alpha_{01}^\prime = \alpha_{01}$ and $\beta_{01}^\prime = \beta_{01}$ or $\alpha_{01}^\prime = \alpha_{01} + N\beta_{01}$ and $\beta_{01}^\prime = -\beta_{01}$. The latter cannot occur due to would-be inconsistencies in the weights. For example, consider weights for the component with mean $\alpha_{01}+\beta_{01}$ under this scenario. On the LHS, the weight would correspond to the probability $\tau_i(c_{01},1;p,q)$, while on the RHS, the weight would correspond to the probability $\tau_i(c_{01},N-1;p^\prime,q^\prime)$. The former quantity changes with $\tilde{d}_{t}$ if holding the total observed degree $\tilde{d}_{t}+\tilde{d}_{nt}$ fixed, since the observed treated degree would affect the probability of a true treated connection, but the latter does not since for very large true degree $d>N_{nt,i}$ we will always have a treated connection. Thus, we have $\alpha_{01}^\prime = \alpha_{01}$ and $\beta_{01}^\prime = \beta_{01}$.

We can then use our identification of the $c_{01}$ components to isolate the remaining, unexplained components, which must correspond to $c_{00}$. If $\beta_{00} \neq 0$, the LHS has $N_{nt,i}+1$ remaining components, while if $\beta_{00} = 0$, the LHS has one component. The same holds for $\beta^\prime_{00}$ and the RHS. Thus, when $\beta_{00} = 0$, $\beta^\prime_{00} = 0$ and we must have $\alpha^\prime_{00} = \alpha_{00}$. On the other hand, if $\beta_{00} \neq 0$, both sides consist of $N_{nt,i} + 1$ components, spaced $\lvert\beta_{00}\rvert$ and $\lvert\beta^\prime_{00}\rvert$ apart respectively. We must have either $\alpha_{00}^\prime = \alpha_{00}$ and $\beta_{00}^\prime = \beta_{00}$ or $\alpha_{00}^\prime = \alpha_{00} + N_{nt,i}\beta_{00}$ and $\beta_{00}^\prime = -\beta_{00}$. Following similar logic as above for the $c_{01}$ components, we can use would-be inconsistencies in the weights to eliminate the second scenario. Namely, consider the weights for the $\alpha_{00}$ component, which is $\tau_i(c_{00},0;p,q)$ for the LHS and $\tau_i(c_{00},N_{nt,i};p^\prime,q^\prime)$ for the RHS. For fixed $\tilde{d}_{t}+\tilde{d}_{nt}$, $\tau_i(c_{00},0;p,q)$ is unaffected by varying $\tilde{d}_{t}$ as all observed connections regardless of treatment status must be falsely observed, while the treatment status of the observed connections will effect the probability of having a treated connection given $N_{nt,i}$ true connections.

\begin{case}
$p > 0$, $q > 0$, and $\beta_{10} = 0$
\end{case}

Next, let us consider the scenario when we have $\beta_{10} = 0$, but $\beta_{00} \neq 0$. For any pair of $\left(\tilde{d}_t,\tilde{d}_{nt}\right)$, the LHS is a normal mixture including $N_{nt,i} + 1$ or $N_{nt,i} + 2$ components with means $\alpha_{01}$ and $\alpha_{00} + \beta_{00}d$ and variance $\sigma^2$ for any $d$ from $\{0,...,N_{nt,i}\}$. Since the number of components does not change for any pair of observed degrees, we have $p^\prime > 0$, $q^\prime > 0$, and $\beta^\prime_{10} = 0$. Following the same logic used in case 1 but reversing the order in which we consider the $c_{00}$ and $c_{01}$ components, we can show $\{\theta_{00},\theta_{01}\} = \{\theta_{00}^\prime,\theta_{01}^\prime\}$.

The alternate scenario involves the case $\beta_{00} = 0$ and $\beta_{01} = 0$. Since we assume there is a non-zero indirect treatment effect ($\theta_{00} \neq \theta_{01}$), the LHS consists of a mixture of two normals with means $\alpha_{00}$ and $\alpha_{01}$. Following the generic identifiability of normal mixtures, the RHS must consist of two normals with the same means. In order for the RHS to have two mixture components regardless of observed degree $\left(\tilde{d}_t,\tilde{d}_{nt}\right)$, we must have $\beta^\prime_{00} = 0$ and $\beta^\prime_{01} = 0$ as well as non-zero mismeasurement in both $p^\prime$ and $q^\prime$. For $q^\prime = 0$, $\tilde{d}_t > 0$ would yield just one mixture component, and similarly with $\tilde{d}_t = 0$ for $p^\prime = 0$. Thus, either $\alpha^\prime_{00} = \alpha_{00}$ and $\alpha^\prime_{01} = \alpha_{01}$ or $\alpha^\prime_{00} = \alpha_{01}$ and $\alpha^\prime_{01} = \alpha_{00}$. If the latter is the case, the weight of the $\alpha_{00}$ component is the probability of no indirect treatment $\sum_d\tau_i(c_{00},d;p,q)$ on the LHS and the probability of indirect treatment $\sum_d\tau_i(c_{01},d;p^\prime,q^\prime)$ on the RHS. These weights must be the same for any pair of $\left(\tilde{d}_t,\tilde{d}_{nt}\right)$. However, when holding $\tilde{d}_t + \tilde{d}_{nt}$ fixed and increasing the number of observed connections to treated individuals $\tilde{d}_t$, the weight of the LHS decreases while the weight of the RHS increase. Thus, we must have $\alpha^\prime_{00} = \alpha_{00}$ and $\alpha^\prime_{01} = \alpha_{01}$.

\begin{case}
$p > 0$ and $q = 0$
\end{case}

First, consider an observation $i$ with at least one observed connection to a treated subject $\tilde{d}_t > 0$. The mixture on the LHS consists of components with means $\alpha_{01} + \beta_{01}d$ corresponding to $\tau_i(c_{01},d;p,q)$ for any $d$ satisfying $d\geq \tilde{d}_{t} + \tilde{d}_{nt}$. If $\beta_{01} = 0$, the LHS will just be one component, while if $\beta_{01} \neq0$, the LHS will have $N-(\tilde{d}_{t} + \tilde{d}_{nt})$ distinct components. 

Suppose for now the latter is true. Then increasing total observed degree $\tilde{d}_{t} + \tilde{d}_{nt}$ decreases the number of components on the LHS. Changing total observed degree has no effect on the number of distinct components when $p,q>0$ or $p=q=0$, while the case $p=0$ and $q>0$ would imply an increase in the number of distinct components. Thus, to match the behavior on the RHS, we must have $p^\prime > 0$ and $q^\prime = 0$. For the components on both sides to have the same set of means, we must have either $\alpha^\prime_{01} = \alpha_{01}$ and $\beta^\prime_{01} = \beta_{01}$ or $\alpha^\prime_{01} = \alpha_{01}+(N-1+\tilde{d}_{t}+\tilde{d}_{nt})\beta_{01})$ and $\beta^\prime_{01} = -\beta_{01}$. We can again invalidate the second case by examining would-be inconsistencies in the weights $\tau_i$, but in this case we can also simply note that the latter scenario cannot be simultaneously valid across multiple choices of $\tilde{d}_{t}+\tilde{d}_{nt}$. Having established $\alpha^\prime_{01} = \alpha_{01}$ and $\beta^\prime_{01} = \beta_{01}$, we can consider observations with $\tilde{d}_t = 0$  and isolate the remaining $\tau_i(c_{00},d;p,q)$ components on the LHS, of which there would be either 1 (if $\beta_{00} = 0$) or $N_{nt,i} - \tilde{d}_{nt} + 1$ (if $\beta_{00} \neq 0$) components. Matching these components on the RHS across multiples values of $\tilde{d}_{nt}$ will avoid the potential case where $\beta^\prime_{00}=-\beta_{00}$ and yield $\alpha^\prime_{00} = \alpha_{00}$ and $\beta^\prime_{00} = \beta_{00}$.

Let us now return to the case where $\beta_{01} = 0$. While we could still find $\beta^\prime_{01} = 0$ and $\alpha^\prime_{01} = \alpha_{01}$, examining the number of components when $\tilde{d}_t > 0$ is not sufficient to imply $p^\prime>0$ and $q^\prime=0$. However, we can attempt to ascertain whether or not this must be the scenario by examining observations with $\tilde{d}_t = 0$. If $\beta_{00} \neq 0$, there would be $N_{nt,i} - \tilde{d}_{nt} + 1$ distinct components on the LHS. A decreasing number of components for these observations as $\tilde{d}_{nt}$ increase is only consistent with $p^\prime>0$ and $q^\prime=0$. From here, we can use the equal spacing of these components as well as the structure imposed by the weights to show $\alpha^\prime_{00} = \alpha_{00}$ and $\beta^\prime_{00} = \beta_{00}$.

Lastly, when both $\beta_{00} = 0$ and $\beta_{01} = 0$, we observe one mixture component with mean $\alpha_{01}$ when $\tilde{d}_{t} > 0$ and two mixture components with means $\alpha_{00}$ and $\alpha_{01}$ when $\tilde{d}_{t} = 0$. Returning to the logic used in the counterpart scenario in case 2, the LHS can only be matched when $p^\prime > 0$ and $q^\prime = 0$. Then the RHS will have one mixture component when $\tilde{d}_{t} > 0$ and two components when $\tilde{d}_{t} == 0$, and we will have $\{\theta_{00},\theta_{01}\} = \{\theta_{00}^\prime,\theta_{01}^\prime\}$.

\begin{case}
$p = 0$ and $q > 0$
\end{case}

This case follows identical logic as case 3 but switching the roles of the $c_{00}$ and $c_{01}$ components. Namely, observations with $\tilde{d}_t = 0$ will isolate the $c_{00}$ components, which in turn can be used to inform observations with $\tilde{d}_t > 0$ to match the $c_{01}$ components.

\begin{case}
$p = 0$ and $q = 0$
\end{case}

For any pair of $\left(\tilde{d}_t,\tilde{d}_{nt}\right)$, the LHS will consist of consist of a single normal distribution. If $p^\prime>0$ or $q^\prime>0$, this behavior could only arise if $\beta_{00} = \beta_{01} = 0$ and $\alpha_{00} = \alpha_{01}$. However, we require $\theta^\prime_{00} \neq \theta^\prime_{11}$, so we must have $p^\prime=0$ and $q^\prime=0$. Observations from two distinct values of $\tilde{d}_t + \tilde{d}_{nt}$ for each of $\tilde{d}_t=0$ and $\tilde{d}_t>0$ will uniquely identify the model parameters $\theta_{00}$ and $\theta_{01}$ respectively.

\pagebreak
\section{Results as tables} \label{section:tables}

\begin{table}[h]
\centering
\begin{tabular}{l|ccc}
Network measure & Intensive Session & Network Intensive & Interaction\tabularnewline
\hline 
\textbf{No mismeasurement} &  &  & \tabularnewline
General measure & 0.205 (0.016) & 0.198 (0.016) & -0.241 (0.023)\tabularnewline
Strong measure & 0.100 (0.012) & 0.120 (0.036) & -0.188 (0.054)\tabularnewline
Weak measure & 0.157 (0.061) & 0.072 (0.044) & -0.095 (0.063)\tabularnewline
 &  &  & \tabularnewline
\textbf{EM method} &  &  & \tabularnewline
General measure & 0.177 (0.025) & 0.229 (0.028) & -0.224 (0.040)\tabularnewline
Strong measure & 0.299 (0.037) & 0.279 (0.040) & -0.577 (0.052)\tabularnewline
Weak measure & 0.160 (0.032) & 0.259 (0.032) & -0.171 (0.048)\tabularnewline
 &  &  & \tabularnewline
\textbf{EM + covariates} &  &  & \tabularnewline
General measure & 0.158 (0.035) & 0.295 (0.039) & -0.322 (0.055)\tabularnewline
Strong measure & 0.177 (0.081) & 0.305 (0.070) & -0.479 (0.116)\tabularnewline
Weak measure & 0.175 (0.095) & 0.293 (0.067) & -0.155 (0.117)\tabularnewline
\end{tabular}
\caption{Differences in insurance knowledge for farmers assigned to second round sessions based on (1) whether they attended an intensive session, (2) whether they had a friend attend a first round intensive session, and (3) the interaction of these two terms. We compare our method to results if we assume there is no mismeasurement in the network for three network measures.}
\label{tab:insurance}
\end{table}

\begin{table}[h]
\centering
\begin{tabular}{l|cc}
Network measure & $p$ & $q$ \tabularnewline
\hline 
General measure & 0.185 & 0.00016 \tabularnewline
Strong measure& 0.872 & 1.23e-06 \tabularnewline
Weak measure&   0.097 & 0.00201 \tabularnewline
\end{tabular}
\caption{Differences in insurance knowledge for farmers assigned to second round sessions based on (1) whether they attended an intensive session, (2) whether they had a friend attend a first round intensive session, and (3) the interaction of these two terms. We compare our method to results if we assume there is no mismeasurement in the network for three network measures.}
\label{tab:pq}
\end{table}

\section{Expanding the number of exposure conditions: Results using the ``linear-in-treated-peers'' model}
\label{sec:LIS}

\begin{figure}[h]
    \centering
    \includegraphics[scale=0.8, width=130mm]{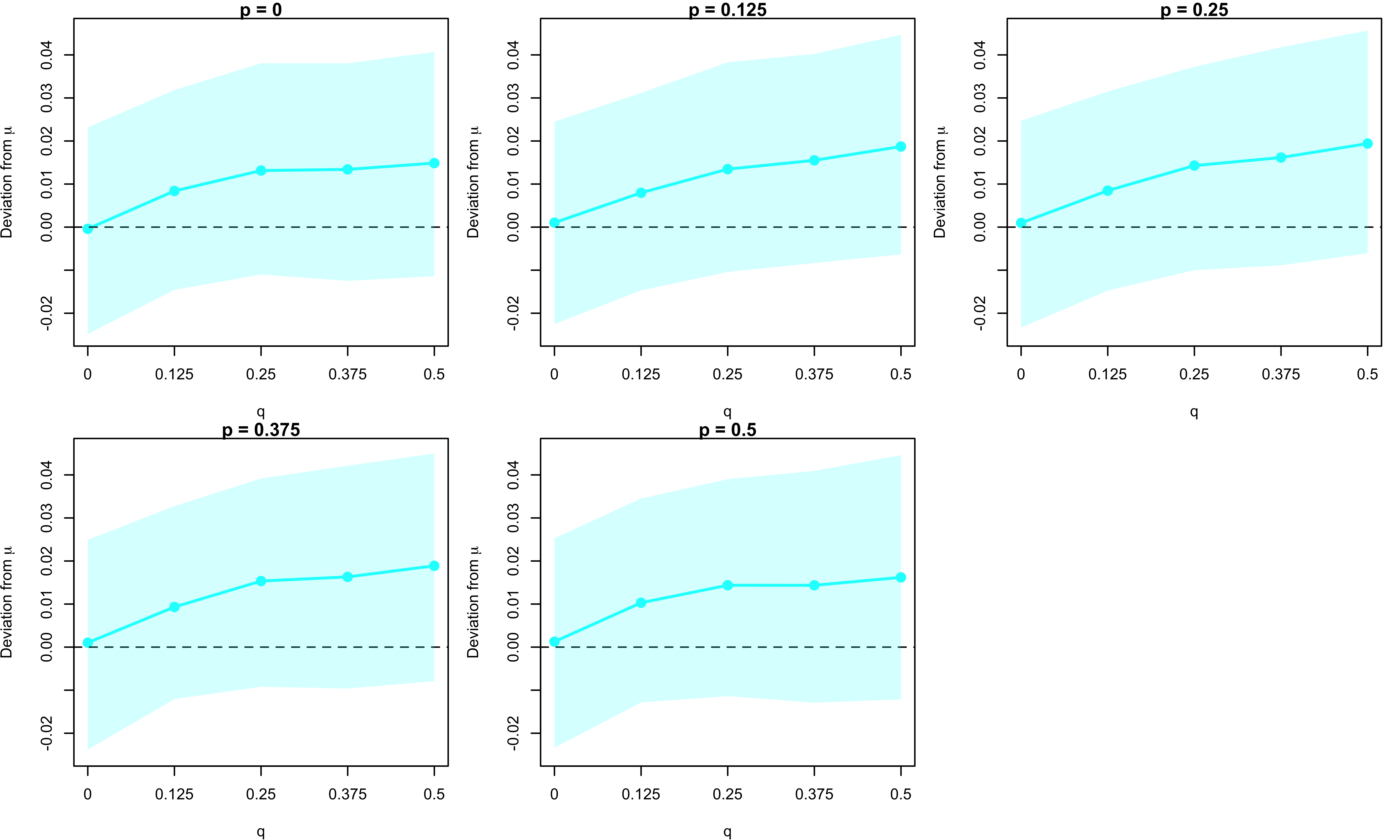}
    \caption{``Linear-in-treated-peers'' estimates of the deviation of the mean outcome across all exposure conditions from their true values under each p value and varying q from 0 to 0.5. The 0.1 and 0.9 quantiles are shaded to give a sense of the variability in these estimates.}
    \label{fig:model1_p}
\end{figure}

\begin{figure}[h]
    \centering
    \includegraphics[scale=1, width=150mm]{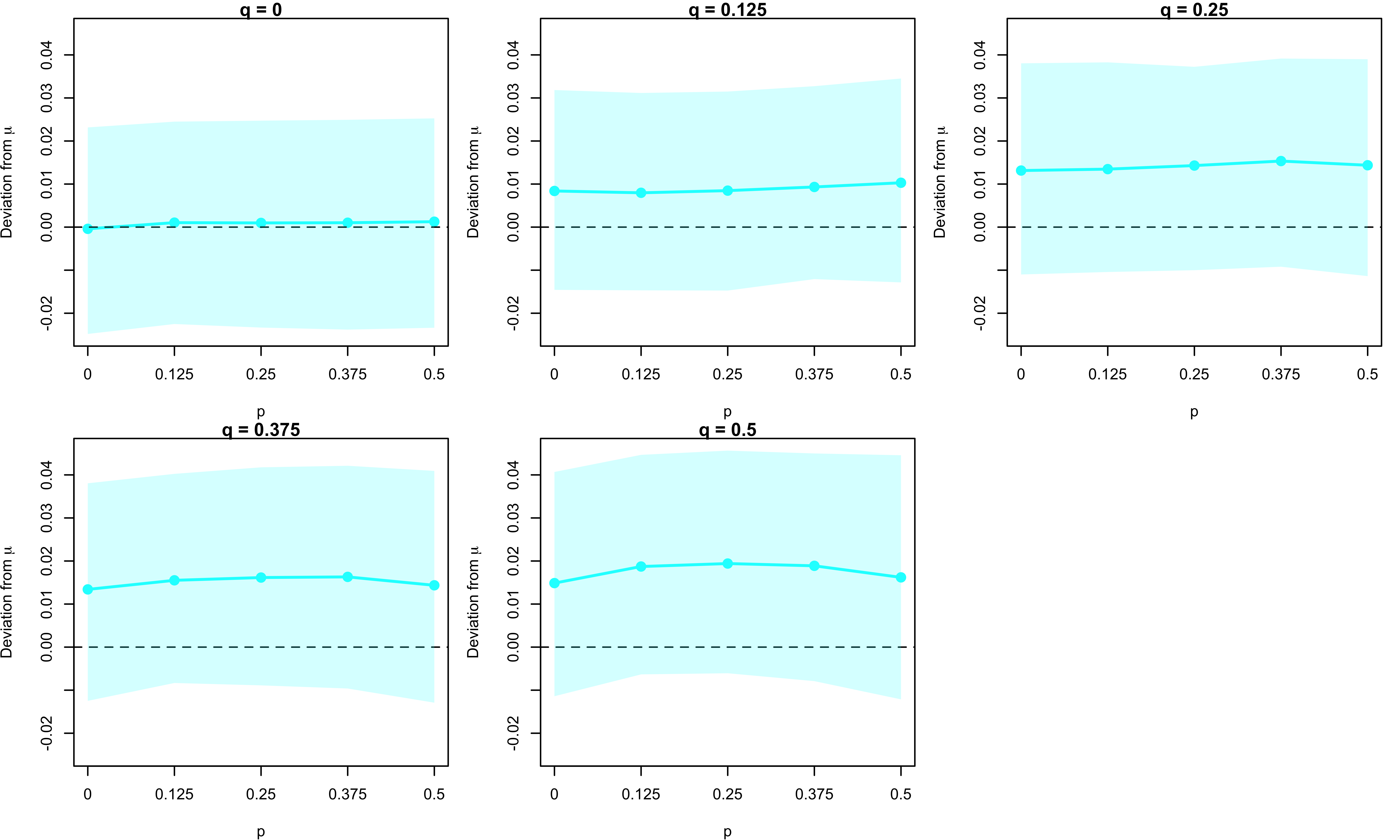}
    \caption{``linear-in-treated-peers'' estimates of the deviation of the mean outcome across all exposure conditions from their true values under each q value and varying p from 0 to 0.5. The 0.1 and 0.9 quantiles are shaded to give a sense of the variability in these estimates.}
    \label{fig:model1_q}
\end{figure}

In this section, we present results where we define exposure conditions and the outcome model in a way that reflects the ``linear-in-treated-peers'' assumptions. First, we simulated networks of size 500 using a Stochastic Block Model (SBM).  For each network, we then simulated the deletion and spurious addition of edges with varying probabilities.  We then fit the ``linear-in-treated-peers'' model by defining exposure conditions such that the maximum exposure condition corresponds to knowing a number of treated peers equal to the maximum observed in-degree. Figures~\ref{fig:model1_p} and~\ref{fig:model1_q} show results for estimating the overall mean outcome across all exposure conditions for various values of $p$ and $q$.  The overall bias remains small (note the axes on the graphs are zoomed-in) across all simulation value parameters.

\begin{figure}[h]
    \centering
        \centering
        \includegraphics[width=.6\textwidth]{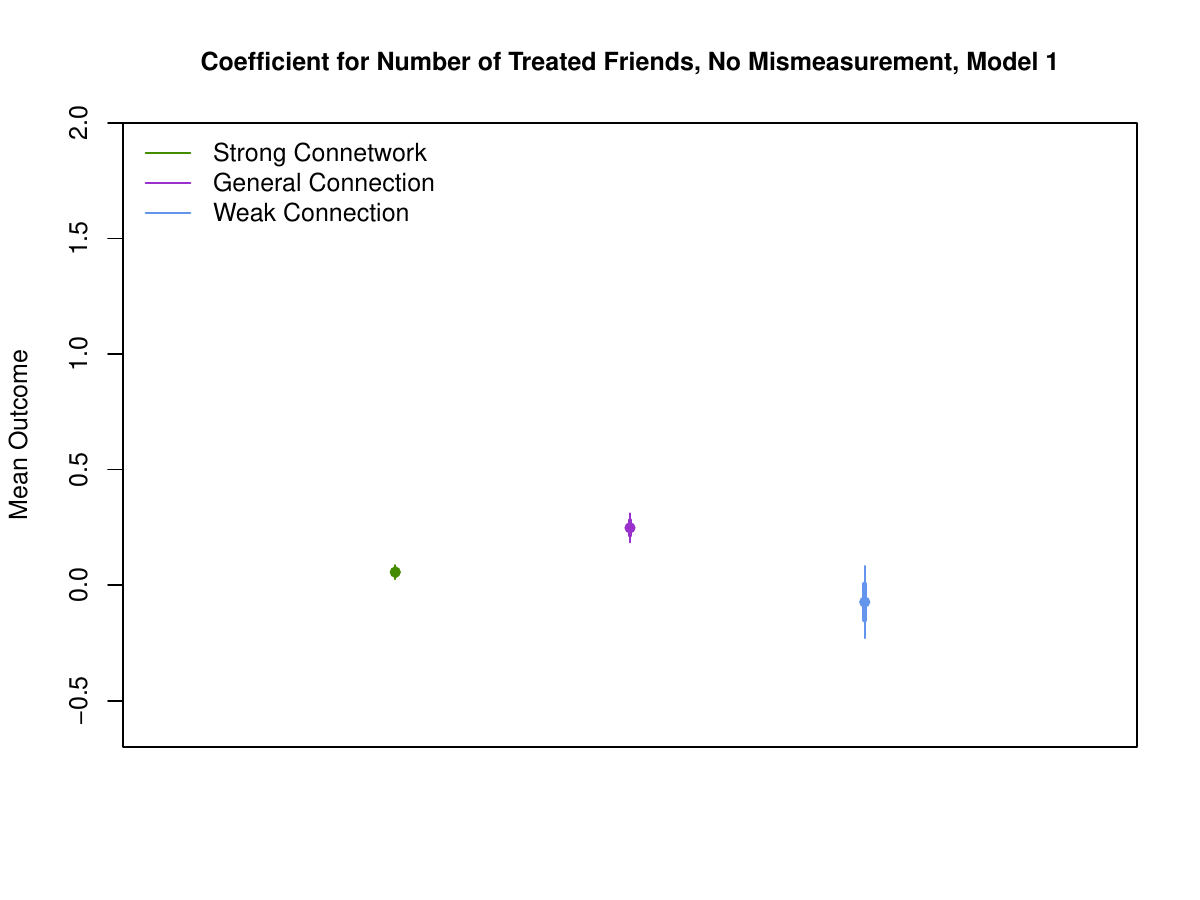}
    \hfill

        \includegraphics[width=.6\textwidth]{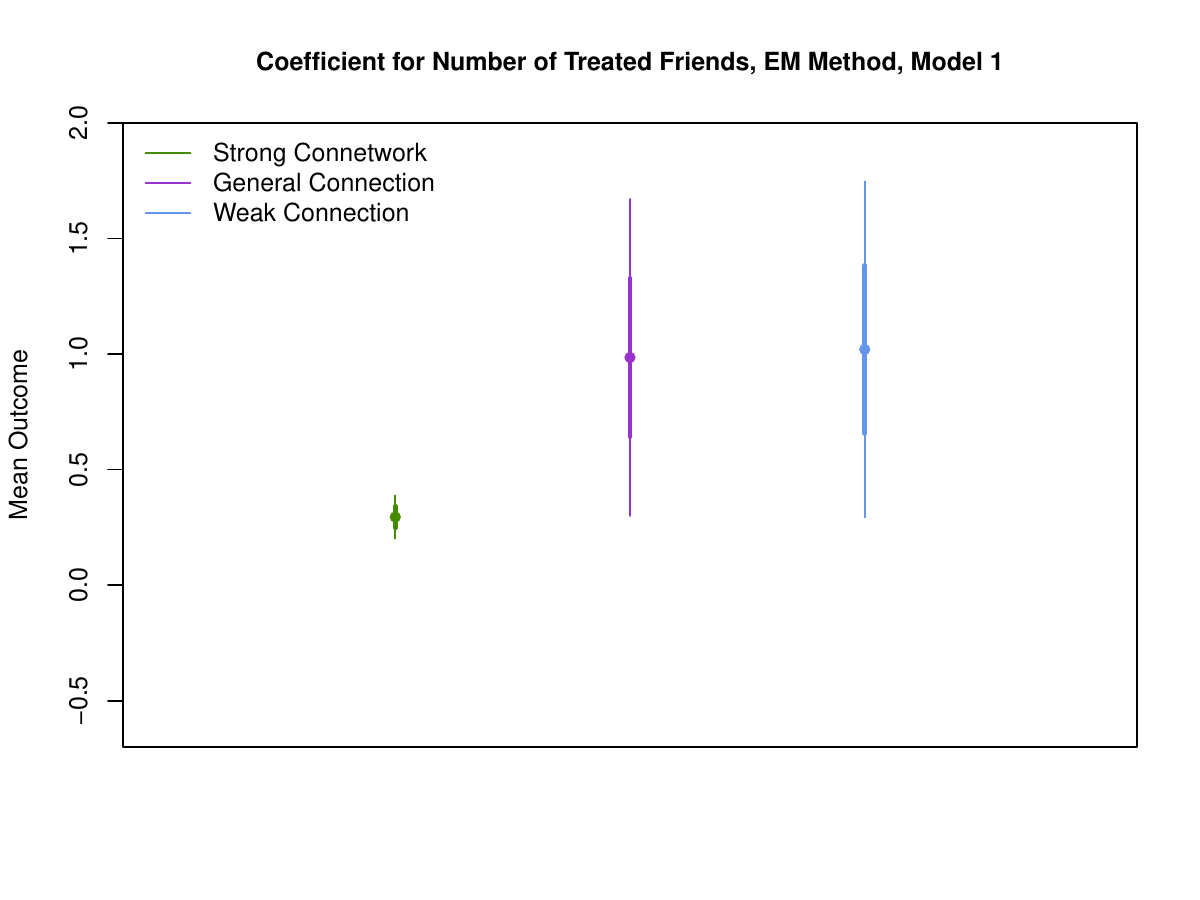}
    \hfill
        \includegraphics[width=.6\textwidth]{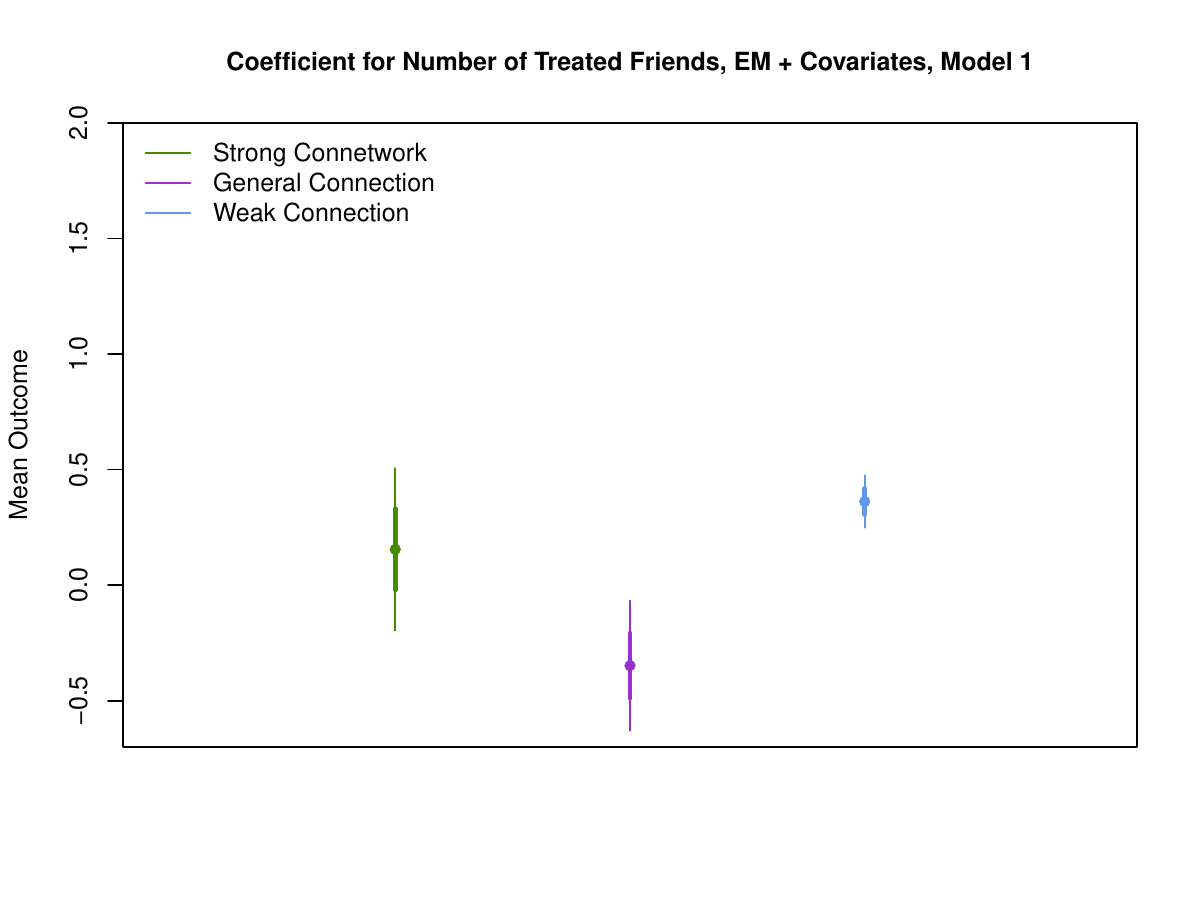}
    \caption{``linear-in-treated-peers'' estimated coefficient for number of treated friends.}
    \label{fig:ins_model1_alpha}
\end{figure}

Next, we applied the model to the insurance information experiment data presented in Section~\ref{section:Application}. Figure \ref{fig:ins_model1_alpha} shows the estimated coefficient for the number of treated peers using the ``linear-in-treated-peers'' specification.  The top panel in the figure shows the results from fitting a GLM and assuming no mismeasurment.  We see that the uncertainty is very low, however, the substantive conclusion would be different depending on the choice of network.  In particular, the weak connection network has an uncertainty interval that overlaps with zero, indicating no peer effects.  The other two networks, however, do not overlap.  When looking at the plot immediately below, however, using the proposed method, all three intervals overall with one-another, but not with zero.  This trend does not hold, however, when adding covariates.  

\end{document}